\documentclass[a4paper,11pt]{article}
\pdfoutput=1 

\usepackage{jcappub} 
\usepackage[toc,page]{appendix}
\usepackage{subcaption}
\usepackage{graphicx}                
\usepackage{multirow}
\usepackage[T1]{fontenc} 
\usepackage[table]{xcolor}
\usepackage{float}
\def\noi{{\noindent}}

\definecolor{verde}{rgb}{0,0.5,0}

\def\be{\begin{equation}}
\def\ee{\end{equation}}
\def\bea{\begin{eqnarray}}
\def\eea{\end{eqnarray}}
\def\be{\begin{equation}}
\def\ee{\end{equation}}
\def\ba{\begin{eqnarray}}
\def\ea{\end{eqnarray}}

\title{\boldmath Small-scale Tests of Inflation}

\author[1,a]{Laura Iacconi,\note{Electronic address: laura.iacconi@port.ac.uk}}
\author[2,a]{Matteo Fasiello,\note{Electronic address: matteo.fasiello@port.ac.uk}}
\author[3,a]{Hooshyar Assadullahi\note{Electronic address: hooshyar.assadullahi@port.ac.uk}} 
\author[4,a]{and David Wands\note{Electronic address: david.wands@port.ac.uk}}
\affiliation{$^{a}$Institute of Cosmology \& Gravitation, University of Portsmouth, PO1 3FX, UK}

\abstract{
We investigate small-scale signatures of the inflationary particle content. We consider the case of a light spin-2 particle sourcing primordial gravitational waves by employing  an effective field theory description. Upon allowing time-dependent sound speeds for the helicity modes, this setup delivers a blue tensor spectrum detectable, for example, by upcoming laser interferometers. Our focus is on the tensor non-Gaussianities that ensue from this field configuration.~After characterising the bispectrum amplitude and shape-function at CMB scales, we move on to smaller scales where anisotropies induced in the tensor power spectrum by long-short modes coupling become the key handle on (squeezed) primordial non-Gaussianities. We identify the parameter space generating percent level anisotropies at scales soon to be probed by SKA and LISA.}

\begin{document}
	\maketitle
	\flushbottom
	
	\section{Introduction}

The inflationary hypothesis, the existence of a period of accelerated expansion in the very early universe, is in exquisite agreement with current observations and has had remarkable success in explaining the origin of structure in the universe. The crucial role inflation plays in early universe dynamics notwithstanding, our understanding of the microphysics of inflation is certainly incomplete. Unanswered questions include the energy scale at which it occurs as well as the identity of the fields that make up the inflationary zoo.  
The simplest viable mechanism for acceleration consists of a single scalar field slowly rolling down its potential. On the other hand,  a richer field content is not just possible but  likely from the top-down  perspective \cite{Baumann:2014nda}. 

In our quest for signatures of the inflationary particle content we will rely on two crucial facts. First, primordial gravitational waves (GW) are a key universal prediction of the inflationary paradigm. Secondly, primordial non-Gaussianities
are the most efficient probe of inflationary interactions. The analysis presented here is centred on the study of a stochastic backgroud of GWs, of primordial origin, that is detectable in the high frequency regime (small scales). In the coming decades, an unprecedented array of experimental missions will perform direct (e.g. Laser Interferometer Space Antenna \cite{Audley:2017drz}, KAGRA \cite{Somiya:2011np}, Einstein Telescope \cite{Maggiore_2020}, DECIGO/BBO \cite{Yagi:2011wg}) and indirect searches (e.g. Simons Observatory \cite{Ade:2018sbj}, LiteBIRD \cite{Hazumi:2019lys}, BICEP Array \cite{Hui:2018cvg}) for the stochastic gravitational waves background (SGWB).

In the single-field slow-roll scenario, GWs display a slightly red-tilted power spectrum\footnote{For an extended analysis of single-field EFT realisations and their observable predictions in the tensor sector see \cite{Capurri:2020qgz}.}, potentially detectable on large scales but unobservable in the foreseeable future at interferometer scales (a possible exception being the proposed ``Big Bang Observer''). It follows that the detection of a primordial signal at small scales would provide very suggestive evidence of a multi-field scenario\footnote{Interesting exceptions exist, such as non-attractor models (see e.g. \cite{Ozsoy:2019slf} for a recent realisation).}. 

 In this work, we explore the observational signatures due to the presence of (extra) spinning fields non-minimally coupled with the inflaton. Particles with spin exhibit an intriguing phenomenology at the level of higher order cosmological correlators, starting with the three-point function (see e.g.\cite{Arkani-Hamed:2015bza}). On the other hand, unitarity constraints severely restrict the allowed mass range for spin $s\geq 2$ fields \cite{Higuchi:1986py,Fasiello:2012rw,Fasiello:2013woa}. Such requirements stem from the notion that particles are unitary irreducible representations of the spacetime isometry group (quasi de Sitter in the case at hand). Given that the inflaton background breaks dS isometries, coupling any additional field content directly to the constant inflaton foliation will weaken the strength of unitarity bounds and effectively allow light spinning particles.

For the purposes of our current study we do not committ to a specific model\footnote{We refer the interested reader to \cite{Biagetti:2017viz,Dimastrogiovanni:2018uqy} for an explicit embedding in the inflationary context of a fully non-linear theory \cite{deRham:2010kj, Hassan:2011zd} comprising a massive spin-2 field.}, opting instead for an effective field theory (EFT) approach and specifically that of \cite{Bordin:2018pca}, where a generalisation of the approach in \cite{Cheung:2007st} has been introduced. Even if the formalism in \cite{Bordin:2018pca} allows for a more general particle content, we focus here on the phenomenology of a spin-2 field, which is likely  the most interesting choice when it comes to GWs observables. For the sourced gravitational wave signal to be the dominant contribution, sub-luminal sound speeds are required. Such a configuration may originate, for example, from  a departure from the adiabatic trajectory in (multi)field space \cite{Achucarro:2012sm}. The original set-up of \cite{Bordin:2018pca} has been extended in \cite{Iacconi:2019vgc} to the case of time-dependent sound speeds for the helicity components of the spin-2 field. This step is necessary to support a blue-tilted GW spectrum, one that is potentially detectable at interferometer scales.

As there are several other realisations that may lead to a sizable GW production on small scales \cite{Bartolo:2016ami}, it is important to further explore the observational consequences of the set-up in \cite{Iacconi:2019vgc} in order to distinguish it from other inflationary mechanisms. In this work we characterise the higher-point statistics of GWs by calculating the tensor 3-point correlation function. The present work goes beyond the analysis performed in \cite{Dimastrogiovanni:2018gkl} in several directions, one being that we are no longer bound by the assumption of a \textit{constant} sub-luminal sound speed. A varying velocity allows for a large GW power spectrum at small scales. The same is true for non-Gaussianities, although a  direct detection of the latter at small scales is general not expected given the suppression of higher-point functions due to propagation effects \cite{Bartolo:2018evs}. 

An interesting case that does not suffer from the same suppression of the signal is that of the \textit{ultra-squeezed} bispectrum. The long mode in this configuration is horizon size (or larger). Two immediate consequences are that (i) the bispectrum cannot be accessed directly given that short modes are e.g. at interferometer scales and the long mode is horizon size; (ii) the long mode and its correlation with two nearly identical short modes is not dampened by propagation effects, much as is the case for the GW power spectrum. The effect of the long wavelength is best probed by the anisotropies it induces on the power spectrum of the two small-wavelength modes \cite{Jeong:2012df, Dai:2013kra, Brahma:2013rua, Dimastrogiovanni:2014ina, Dimastrogiovanni:2015pla}. This configuration has been recently studied in \cite{Dimastrogiovanni:2019bfl}: a primordial ultra-squeezed tensor bispectrum induces a quadrupolar modulation on the corresponding power spectrum. In this context, anisotropies represents our best handle on inflationary GW interactions. 
\noi{In this work we calculate the tensor bispectrum contributions mediated by a spin-2 field. We study the bispectrum amplitude and shape-function in different regimes. The main focus is on the case of scale-dependent sound speeds for the helicity modes, a configuration whose parameter space we constrain by employing observational bounds from the CMB. In the high-frequencies regime, we explore the ability of SKA and LISA to indirectly probe non-Gaussianities in the  ultra-squeezed configuration by testing anisotropies of the GW power spectrum. We find that, if GW detectors are able to discern percent level quadrupolar anisotropies of the GW spectrum, this will enable us to rule out (in) large portions of the effective theory parameter space.}\\
This paper is organised as follows. In Section \ref{sec:theory} we review the EFT set-up and the results on the GW spectrum  that will be our starting points.  In Section \ref{sec:calculation} we calculate the tensor bispectrum mediated by light spin-2 fields and study its amplitude and shape-function. In Section \ref{sec:CMB bounds} we focus on the GW observables at large scales. In Section \ref{sec:detectability} we instead focus on small scales and show how the squeezed bispectrum may be tested in this regime. We summarise our findings and point to future research in Section \ref{sec:conclusions}. Details of the bispectrum calculation may be found in the Appendices. 
\medskip
\\ \textit{Conventions:} The spin-2 tensor modes are expanded in  Fourier components as $\hat{\mathcal{T}}_{ij}(\mathbf{x}, \tau)=\int \frac{d^3 k}{(2\pi)^3} \, \mbox{e}^{i \, \mathbf{k}\cdot\mathbf{x}}\, \hat{\mathcal{T}}_{\mathbf{k},\, ij}(\tau)$, where $\tau$ is conformal time ($d\tau=dt/a$) and $\hat{\mathcal{T}}_{ij}(\mathbf{x}, \tau)$ is a place holder for the tensor metric perturbation $\hat{\gamma}_{ij}(\mathbf{x}, \tau)$ and the extra spin-2 field $\hat{\sigma}_{ij}(\mathbf{x}, \tau)$. The modes are decomposed by means of the transverse and traceless polarization tensors $\epsilon_{ij}^\lambda(\hat{\mathbf{k}})$ as  $\hat{\mathcal{T}}_{\mathbf{k},\, ij}(\tau)=\sum_{\lambda=L,R}\epsilon_{ij}^\lambda(\hat{\mathbf{k}})\, \hat{\mathcal{T}}^\lambda_\mathbf{k}(\tau)$, where $\hat{\mathcal{T}}^\lambda_\mathbf{k}(\tau)=\hat{a}^\lambda_\mathbf{k} \mathcal{T}^\lambda_k (\tau)+ \hat{a}^{\lambda \, \dagger}_\mathbf{-k} \mathcal{T}^{\lambda *}_k(\tau)$. The creation and annihilation operators satisfy $[\hat{a}^\lambda_\mathbf{k}, \hat{a}_\mathbf{k'}^{\lambda ' \,\dagger} ]=(2\pi)^3\,\delta^{\lambda \lambda'}\,\delta^{(3)}(\mathbf{k}-\mathbf{k'})$ and $\mathcal{T}^\lambda_k(\tau)$ is the mode function.

\section{Review of the inflationary set-up}
\label{sec:theory}
\begin{figure}
\centering
\includegraphics[scale=0.5]{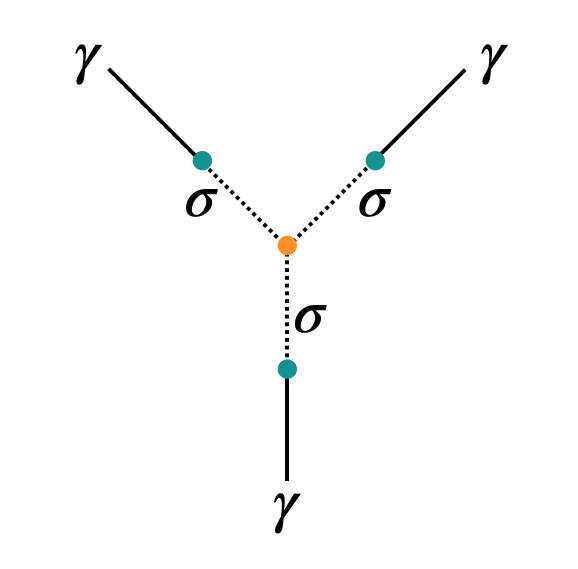}
\caption{Diagrammatic contribution to the tensor bispectrum mediated by a light spin-2 field. The vertices making up the diagram correspond to the quadratic interaction $\mathcal{L}^{(2)} \sim \rho \, \sigma^{ij}\dot{\gamma}_{ij}$ (green) and the cubic self-interaction $\mathcal{L}^{(3)} \sim \mu  (\sigma_{ij})^3$ (orange).}
\label{fig:diagram}
\end{figure}

Let us briefly introduce our starting point, namely the operators in the EFT Lagrangian of \cite{Bordin:2018pca} elucidating the dynamics of the spin-2 field and its coupling with the curvature and tensor fluctuations. At quadratic order the Lagrangian for $\sigma_{ij}(\mathbf{x},t )$ reads 
\begin{equation}
\label{quadratic lagrangian}
\begin{split}
     \mathcal{L}^{(2)}=&\; \frac{1}{4} a^3 \Big[(\dot{\sigma}^{ij})^2 -c_2^2a^{-2}(\partial_i \sigma^{jk})^2 -\frac{3}{2}(c_0^2-c_2^2)a^{-2}(\partial_i \sigma^{ij})^2 -m_\sigma^2(\sigma^{ij})^2 \Big]+ \\
     & + a^3 \Big[- \frac{\rho}{\sqrt{2\epsilon}H}\, a^{-2} \, \partial_i \partial_j \pi \sigma^{ij}  + \frac{\rho}{2}  \, \dot{\gamma}_{ij} \sigma^{ij} \Big]\;,
\end{split}
\end{equation}
where the free Lagrangian is spelled out in the first line, whereas the second line includes the interaction terms with the metric perturbations $\zeta(\mathbf{x},t )= - H \pi(\mathbf{x},t )$ and $\gamma_{ij}(\mathbf{x},t )$. The quantity $a$ is the scale factor, $H \equiv \dot{a}/a$ is the Hubble rate during inflation and $c_i$ is the sound speed of the corresponding helicity component of the spin-2 field.
\\ To ensure that the interaction Lagrangian can be treated perturbatively and to avoid gradient instabilities, the coupling must satisfy $\rho/H\ll \sqrt{\epsilon c_0^2}$ (see \cite{Bordin:2018pca}), where $\epsilon=-\dot{H}/H^2$ is the standard slow-roll parameter. {Such bound also defines the weak-mixing regime for the spin-2 field, where the mode function of the $i$th-helicity component is well-described by the solution to the free-field equation,}
\begin{equation}
\label{mode function sigma}
  \sigma_k(\tau)= \sqrt{\frac{\pi}{2}}H (-\tau)^{3/2}\, H_\nu^{(1)}(-c_i k\tau)  \;,
\end{equation}
with $H_\nu^{(1)}$ the Hankel function of the first kind. 
 The quadratic interactions in Eq.~\eqref{quadratic lagrangian} couple the helicity-0 component of the spin-2 field with the scalar metric perturbation and the helicity-2 with the tensor perturbation. As a result, the field $\sigma$  sources both scalar and tensor power spectra, to obtain:
\begin{gather}
\label{scalar power spectrum}
\mathcal{P}_\zeta(k)=\frac{ H^2}{8 \pi^2 M_{Pl}^2 \epsilon}\Big[1+ \frac{\mathcal{C}_\zeta(\nu)}{\epsilon c_0^{2\nu}} \Big(\frac{\rho}{H} \Big)^2 \Big] \;, \\
\label{tensor power spectrum}
    \mathcal{P}_\gamma(k)=\frac{2 H^2}{\pi^2 M_{Pl}^2}\Big[1+ \frac{\mathcal{C}_\gamma(\nu)}{c_2^{2\nu}} \Big(\frac{\rho}{H} \Big)^2 \Big] \;,
\end{gather}
where in both expressions the first term is due to vacuum fluctuations whereas the second is sourced by the spin-2 field. The quantity $\nu$ is given by $\nu=\sqrt{9/4-(m_{\sigma}/H)^2}$  and the functions $\mathcal{C}_\gamma(\nu)$ and $\mathcal{C}_\zeta(\nu)$ can be computed analytically and are typically of $\mathcal{O}(1)-\mathcal{O}(100)$ \cite{Bordin:2018pca}. As shown in \cite{Iacconi:2019vgc}, there are phenomenologically interesting ansatze according to which one can safely assume that the scalar power spectrum is dominated by the vacuum contribution across all scales of interest. The case of time-dependent sound speed for the spin-2 helicity components has also been explored in \cite{Iacconi:2019vgc}. There, as well as in this work, we will employ the related expression for the sound speed as a function of $k$:
\begin{equation}
\label{c2(k)}
    c_2(k)=c_{2\, in} \Big(\frac{k}{a_0 H_0} \Big)^{s_2} \;, 
\end{equation}
where $s_2 \equiv \dot{c}_2 / (H c_2) $ is assumed constant for simplicity and we take the size of the universe today as the pivot scale (one could alternatively use $k^*=k^{\rm }_{\rm CMB}$, such as is done in \cite{Bartolo:2016ami}), i.e. the scale where $c_2(k^*)= c_{2\, in}$. Such $k$ dependence is obtained by virtue of the fact that cosmological correlators give the leading contribution at horizon crossing. At the horizon a precise relation is in place between wavenumber and conformal time, for example $|k\tau|\simeq 1$  for the tensor fluctuations $\gamma_{ij}$. The sound speed is assumed to be slowly varying ($|s_2|\ll 1$) and, as a result, the next-to-leading corrections to the mode function in Eq.\eqref{mode function sigma} can be safely neglected \cite{Chen_2007}. The resulting scaling of the tensor power spectrum is given by  
\begin{figure}
\centering
\captionsetup[subfigure]{justification=centering}
   \begin{subfigure}[b]{0.49\textwidth}
    \includegraphics[width=\textwidth]{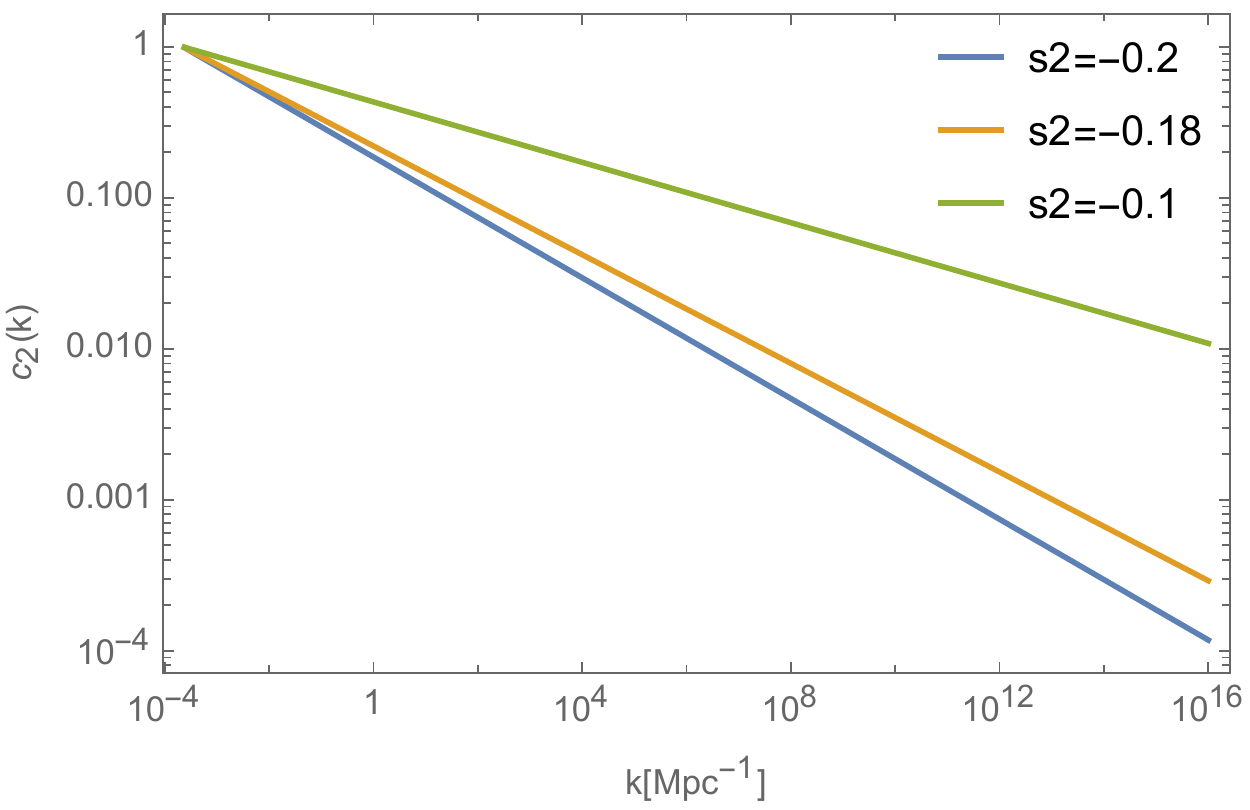}

  \end{subfigure}
  \begin{subfigure}[b]{0.49\textwidth}
    \includegraphics[width=\textwidth]{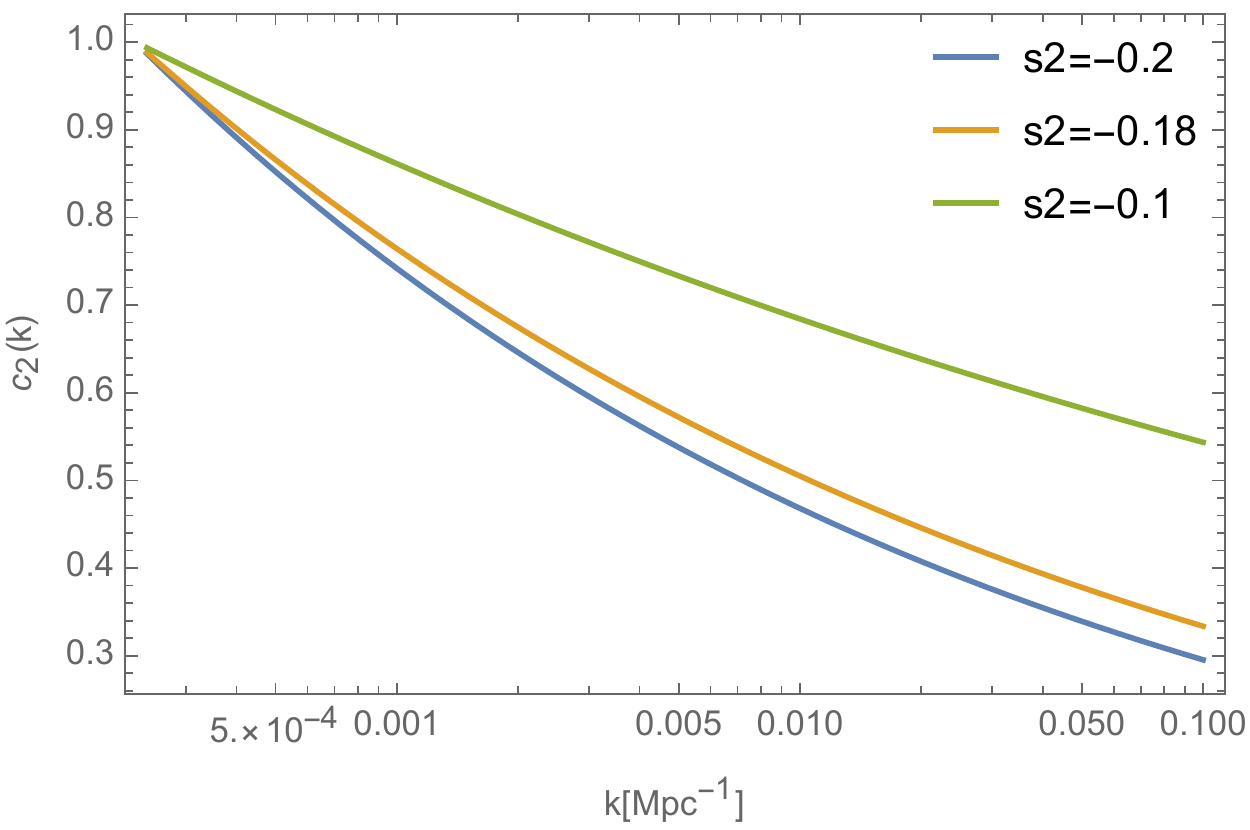}

  \end{subfigure}
  \caption{Working example for the evolution of $c_2(k)$. In both panels the function \eqref{c2(k)} is plotted, with $c_{2 \, in}=1$ and different lines representing different choices of $s_2<0$. On the left, the evolution of $c_2(k)$ is shown over a range of scales which spans from the size of the observable horizon $a_0H_0$ to LIGO scales. On the right, the focus is on the large scale behavior of $c_2(k)$.}
  \label{fig:c2(k)eq}
\end{figure}
\begin{equation}
\label{tilt in tensor explicit}
    \mathcal{P}_\gamma(k) \propto \frac{1}{{c_{2\,in}}^{2\nu}} \Big(\frac{k}{a_0 H_0} \Big)^{-2\nu s_2}.
\end{equation}
{For a decreasing sound speed $(s_2<0)$ and an appropriate choice of the other parameters, the GW signal is detectable at interferometer scales by upcoming probes, including LISA. One such configuration corresponds to the parameters }
\begin{equation}
\label{interesting configuration}
    \{H=6.1\times 10^{13}\mbox{GeV}, \, \nu=1.4, \, c_{2\, in}=1\} \;.
\end{equation}
We stress that this is just one of point in an entire region of parameter space that would generate a detectable signal. In Fig.~\ref{fig:c2(k)eq}, the function \eqref{c2(k)} is plotted with initial condition $c_{2\, in}=1$ for three different values of $s_2$. In particular, an upper bound $|s_2|_{max}$ is identified to ensure we stay within the perturbative regime  \cite{Iacconi:2019vgc}. On the left panel the evolution over a large range of scales is displayed, while in the right panel the focus is on the large scale behavior. The effective theory Lagrangian also comprises cubic self-interactions for the $\sigma$ field,
\begin{equation}
\label{L3}
    \mathcal{L}^{(3)} =-a^3  \mu  (\sigma_{ij})^3 \;,
\end{equation}
where $\mu/H\ll1$ to ensure perturbativity. As pointed out in \cite{Dimastrogiovanni:2018gkl}, the structure of the interaction sector of the theory closely resembles the one in quasi-single field inflation \cite{Chen:2009zp}. In particular, the 3-point correlation function of tensor perturbations receives a contribution mediated by the light spin-2 field, as  shown in Fig.~\ref{fig:diagram}.  In Section  \ref{sec:calculation} we shall investigate the tensor bispectrum, its amplitude and shape dependence. 

\section{Tensor bispectrum}
\label{sec:calculation}
A key observable when it comes to testing inflationary interactions, (tensor) non-Gaussianities are typically more constrained at CMB scales (e.g. by data from the Planck mission) than in the complementary high-frequency regime. With the advent of new, more sensitive, GW probes we can aim also at testing those inflationary scenarios that support a large signal at small scales. The set-up we are considering here is one such example and the EFT approach we adopt is the ideal framework to expand our analysis towards an ever richer particle spectrum. Our current focus is on an extra spin-2 field $\sigma$, directly coupled with the standard tensor degrees of freedom field and mediating their interactions. We organise the various contributions to the tensor 3-point correlation function in the following fashion 
\begin{equation}
\label{bispectrum def}
    \langle \gamma_{\mathbf{k_1}}^{\lambda_1}\gamma_{\mathbf{k_2}}^{\lambda_2}\gamma_{\mathbf{k_3}}^{\lambda_3}\rangle = (2\pi)^3 \delta^{(3)}(\mathbf{k_1+k_2+k_3}) \,\mathcal{A}^{\lambda_1 \lambda_2 \lambda_3}\, B_\sigma (k_1,k_2,k_3) \;,
\end{equation}
where the function $\mathcal{A}^{\lambda_1 \lambda_2 \lambda_3}$ accounts for the different polarizations. The quantity $B_\sigma$ is given by
\begin{equation}
\label{Bispectrum all terms}
    B_\sigma (k_1,k_2,k_3)= \frac{12 \pi^3}{k_1^4 \, k_2 \, k_3} \frac{\mu}{H} \Big(\frac{\rho}{M_{Pl}} \Big)^3 \Big[ \mathcal{M}_A + \mathcal{M}_B + \mathcal{M}_C \Big] + \mbox{5 perms} \;, 
\end{equation}
where 
\begin{equation}
\label{FA}
\begin{split}
 \mathcal{M}_A & (\nu, k_1, k_2, k_3) = \int_{-\infty}^{0} dx_1 \;\int_{-\infty}^{x_1} dx_2 \;\int_{-\infty}^{x_2} dx_3 \;\int_{-\infty}^{x_3} dx_4  \sqrt{\frac{x_2}{x_1 x_3 x_4}}\; \sin{(-x_1)} \\
 & \Im\Big[H_\nu^{(1)}(-c_2(k_1) x_1) H_\nu^{(2)}(-c_2(k_1) x_2)\Big] \;
 \Im \Big[\mbox{e}^{-i k_3/k_1 x_4} H_\nu^{(1)}(-c_2(k_3) \frac{k_3}{k_1} x_4) H_\nu^{(2)}(-c_2(k_3) \frac{k_3}{k_1} x_2)\Big] \\
 & \Im\Big[ \mbox{e}^{i k_2/k_1 x_3 } H_\nu^{(1)}(-c_2(k_2) \frac{k_2}{k_1} x_2) H_\nu^{(2)}(-c_2(k_2) \frac{k_2}{k_1} x_3)\Big] \;, 
\end{split}
\end{equation}

\begin{equation}
\label{FB}
\begin{split}
 \mathcal{M}_B & (\nu, k_1, k_2, k_3) =\int_{-\infty}^{0} dx_1 \;\int_{-\infty}^{x_1} dx_2 \;\int_{-\infty}^{x_2} dx_3 \;\int_{-\infty}^{x_3} dx_4  \sqrt{\frac{x_3}{x_1 x_2 x_4}}\; \sin{(-x_1)}\, \sin{(- \frac{k_2}{k_1} x_2)} \\
 & \Im\Big[H_\nu^{(1)}(-c_2(k_1) x_3) H_\nu^{(1)}(-c_2(k_2)\frac{k_2}{k_1} x_3) H_\nu^{(2)}(-c_2(k_1)  x_1) H_\nu^{(2)}(-c_2(k_2) \frac{k_2}{k_1} x_2) \Big] \\
&  \Im \Big[\mbox{e}^{-i k_3/k_1 x_4} H_\nu^{(1)}(-c_2(k_3) \frac{k_3}{k_1} x_4) H_\nu^{(2)}(-c_2(k_3) \frac{k_3}{k_1} x_3)\Big] \;, 
\end{split}
\end{equation}
\begin{equation}
\label{FC}
\begin{split}
 \mathcal{M}_C & (\nu, k_1, k_2, k_3) = \int_{-\infty}^{0} dx_1 \;\int_{-\infty}^{x_1} dx_2 \;\int_{-\infty}^{x_2} dx_3 \;\int_{-\infty}^{x_3} dx_4  \sqrt{\frac{x_4}{x_1 x_2 x_3}}\; \sin{(-x_1)}\, \sin{(- \frac{k_2}{k_1} x_2)}\\
 & \sin{(- \frac{k_3}{k_1} x_3)}\; \Im\Big[H_\nu^{(1)}(-c_2(k_1) x_4) H_\nu^{(1)}(-c_2(k_2)\frac{k_2}{k_1} x_4) H_\nu^{(1)}(-c_2(k_3) \frac{k_3}{k_1} x_4) \\
 & H_\nu^{(2)}(-c_2(k_1) x_1)H_\nu^{(2)}(-c_2(k_2) \frac{k_2}{k_1} x_2)H_\nu^{(2)}(-c_2(k_3) \frac{k_3}{k_1} x_3)\Big] \;,
\end{split}
\end{equation}
and $c_2(k)$ is given in Eq.\eqref{c2(k)}.
The structure of the integrals is due to the use of the nested commutator form in the in-in formalism computation. The dimensionless integration variables are defined as $x_i \equiv k_1 \tau_i$. Let us now focus on the bispectrum in two specific limits, the equilateral and ``local'' one.

\subsection{Equilateral configuration}
In the equilateral configuration ($k_1=k_2=k_3\equiv k$) the bispectrum reads 
\begin{equation}
\label{Bsigma equilateral}
     B_{(\sigma) eq}(k)= \frac{72 \pi^3}{k^6} \frac{\mu}{H} \Big(\frac{\rho}{M_{Pl}} \Big)^3 s_{eq}(\nu,k) \;, 
\end{equation}
where 
\begin{equation}
\label{seq}
\begin{split}
    s_{eq}&(\nu,k)=\int_{-\infty}^{0} dx_1 \;\int_{-\infty}^{x_1} dx_2 \;\int_{-\infty}^{x_2} dx_3 \;\int_{-\infty}^{x_3} dx_4 \; \Big\{
    \sqrt{\frac{x_2}{x_1 x_3 x_4}}\; \sin{(-x_1)} \times \\
    &\Im\Big[H_\nu^{(1)}(-c_2(k) x_1) H_\nu^{(2)}(-c_2(k) x_2)\Big] \;
 \Im \Big[\mbox{e}^{-i x_4} H_\nu^{(1)}(-c_2(k) x_4) H_\nu^{(2)}(-c_2(k)  x_2)\Big]\times \\
 &\Im\Big[ \mbox{e}^{i x_3 } H_\nu^{(1)}(-c_2(k) x_2) H_\nu^{(2)}(-c_2(k)  x_3)\Big] +\sqrt{\frac{x_3}{x_1 x_2 x_4}}\; \sin{(-x_1)}\, \sin{(- x_2)}  \times  \\
 &\Im\Big[H_\nu^{(1)}(-c_2(k) x_3) H_\nu^{(1)}(-c_2(k) x_3) H_\nu^{(2)}(-c_2(k)  x_1) H_\nu^{(2)}(-c_2(k)  x_2) \Big] \times \\
 &\Im \Big[\mbox{e}^{-i x_4} H_\nu^{(1)}(-c_2(k) x_4) H_\nu^{(2)}(-c_2(k) x_3)\Big] + \sqrt{\frac{x_4}{x_1 x_2 x_3}}\; \sin{(-x_1)}\, \sin{(- x_2)} \sin{(-  x_3)}\times \\
 &\Im\Big[H_\nu^{(1)}(-c_2(k) x_4) H_\nu^{(1)}(-c_2(k) x_4) H_\nu^{(1)}(-c_2(k) x_4) H_\nu^{(2)}(-c_2(k) x_1)H_\nu^{(2)}(-c_2(k)  x_2)H_\nu^{(2)}(-c_2(k) x_3)\Big] \Big\}\;.
    \end{split} 
\end{equation}
The integrals in Eq.\eqref{seq} need to be evaluated numerically.
\begin{figure}
\centering
\captionsetup[subfigure]{justification=centering}
   \begin{subfigure}[b]{0.49\textwidth}
    \includegraphics[width=\textwidth]{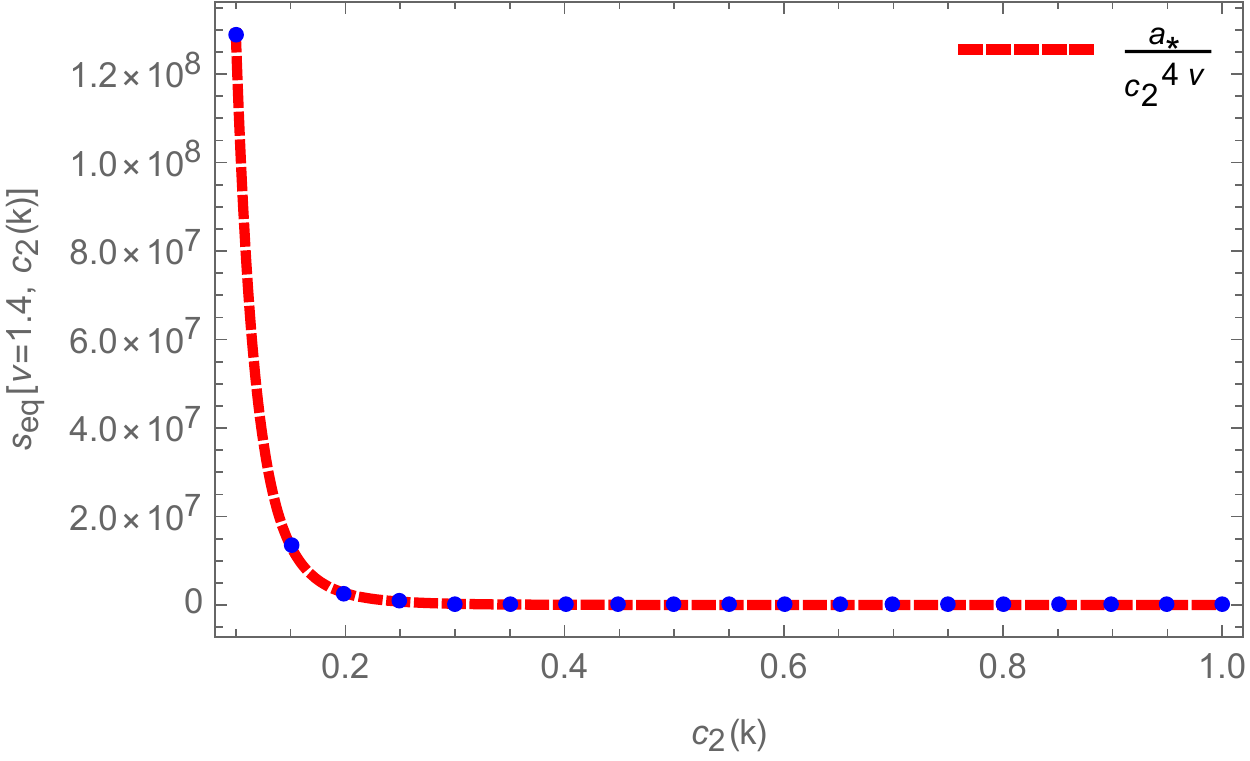}

  \end{subfigure}
  \begin{subfigure}[b]{0.49\textwidth}
    \includegraphics[width=\textwidth]{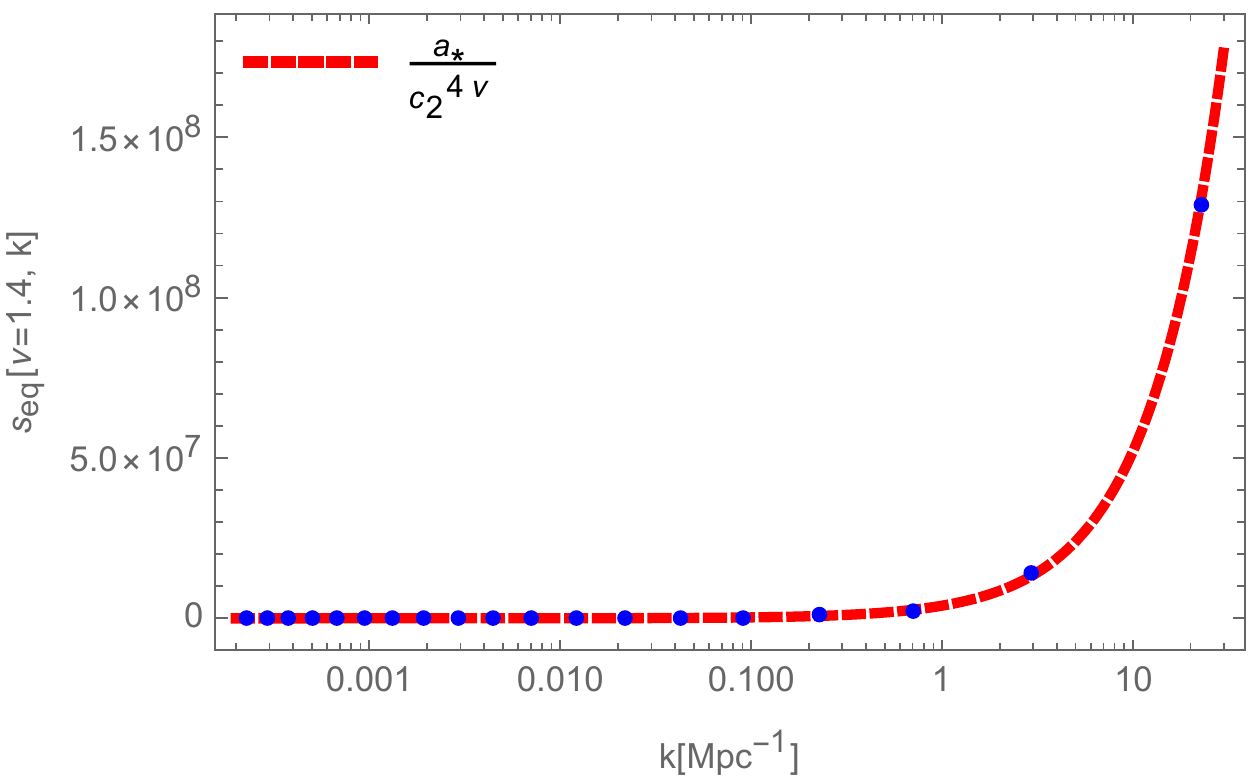}

  \end{subfigure}
  \caption{Results for $s_{eq}(\nu=1.4)$. On the left panel we represent the results as a function of $c_2(k)$, keeping the $k$-dependence implicit, whereas on the right we replace Eq.\eqref{c2(k)} and make explicit the dependence on the scale. In both plots, dots represent numerical results and the dotted red line the fitting functions \eqref{fit 1.4 eq} (left) and \eqref{fit 1.4 eq k} (right).}
  \label{fig:seq14}
\end{figure}
In Fig.~\ref{fig:seq14}, blue dots represent the numerical values of Eq.\eqref{seq} computed for $\nu=1.4$, which corresponds to $m\simeq0.54 H$. As expected, $s_{eq}$ increases for small values of the sound speed, enhancing the resulting bispectrum. The numerical results are fitted with a power law 
\begin{equation}
\label{ansatz power law}
 s_{eq} [\nu, c_2(k)]= \frac{a_\star}{c_2(k)^{4\nu}}   \;.
\end{equation}
The validity of the approximation with a power law is, of course, not surprising considering the usual scaling   $B_{(\sigma)}(k) \propto F_{\rm NL}\, P_\gamma(k)^2$.  For $\nu=1.4$, the fit produces
\begin{equation}
\label{fit 1.4 eq}
    s_{eq}[\nu=1.4, \, c_2(k)] \simeq \frac{324.4}{c_2(k)^{5.6}} \;,
\end{equation}
which is plotted on the left panel of Fig.~\ref{fig:seq14}. One can write explicitly the $k$-dependence, to obtain \begin{equation}
\label{fit 1.4 eq k}
  s_{eq}[\nu=1.4, \, k]\simeq 324.4 \Big(\frac{k}{a_0 H_0}\Big)^{-5.6 s_2} \;,
\end{equation}
as displayed in the right panel of Fig.~\ref{fig:seq14} for $s_2=-0.2$.
\begin{figure}
\centering
\includegraphics[scale=0.7]{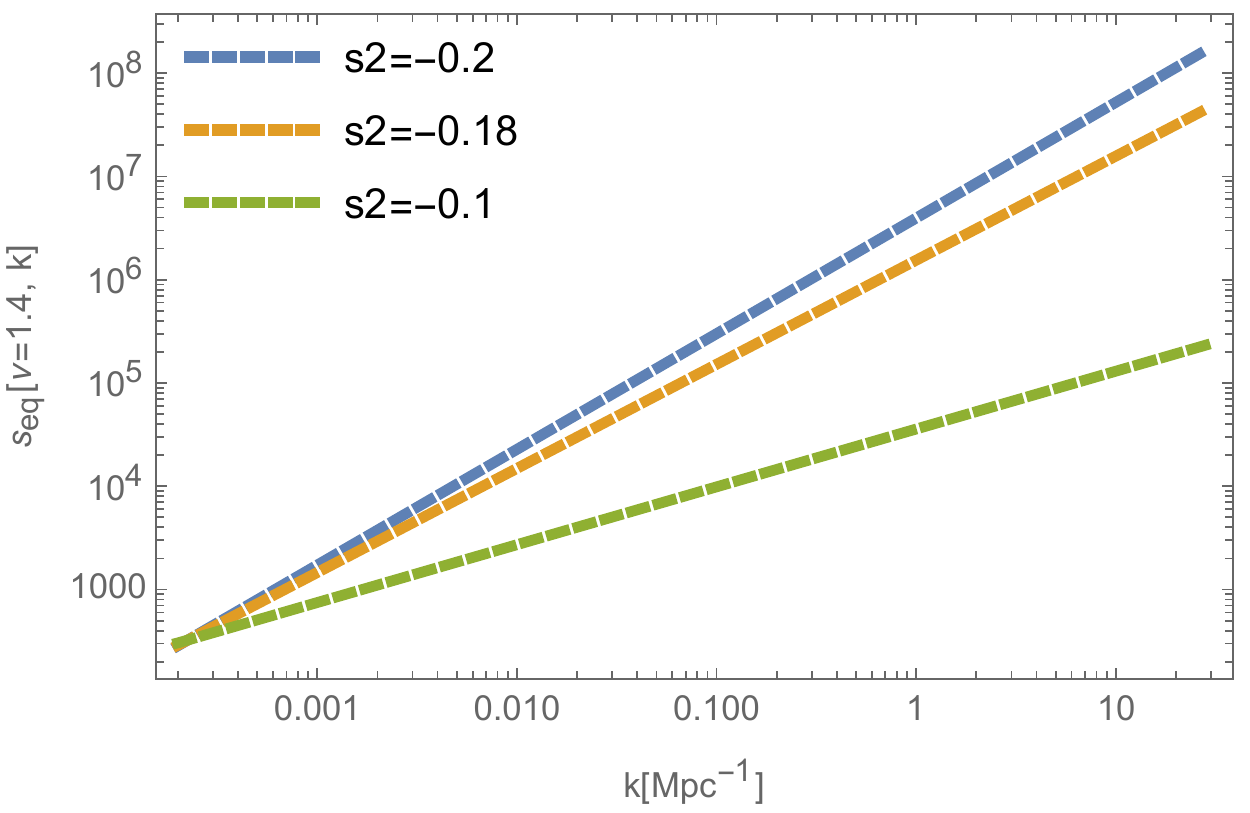}
\caption{Investigating the effect of $s_2$ on $s_{eq}(\nu=1.4, \, k)$. The larger $|s_2|$ is, the faster the sound speed decreases (see Fig.~\ref{fig:c2(k)eq}), amplifying the magnitude of the sourced bispectrum at a given scale.}
\label{fig:s2 effect}
\end{figure}
The value of $s_{eq}$  increases on small scales as the sound speed $c_2$ decreases. In Fig.~\ref{fig:s2 effect}, the fit in \eqref{fit 1.4 eq k} is shown for different values of $s_2$. 
Similar plots for different mass values, $\nu=\{ 0.4,\, 0.8,\, 1.1, \, 1.48\}$ are included in Appendix \ref{app seq}. Our analysis shows that the lighter the spin-2 is, the greater is the size of $s_{eq}$. This is intuitively clear given the  suppression effect of a heavy mass on cosmological correlators.  We shall now consider the squeezed limit.

\subsection{Squeezed configuration}
\begin{figure}
\centering
\includegraphics[scale=0.6]{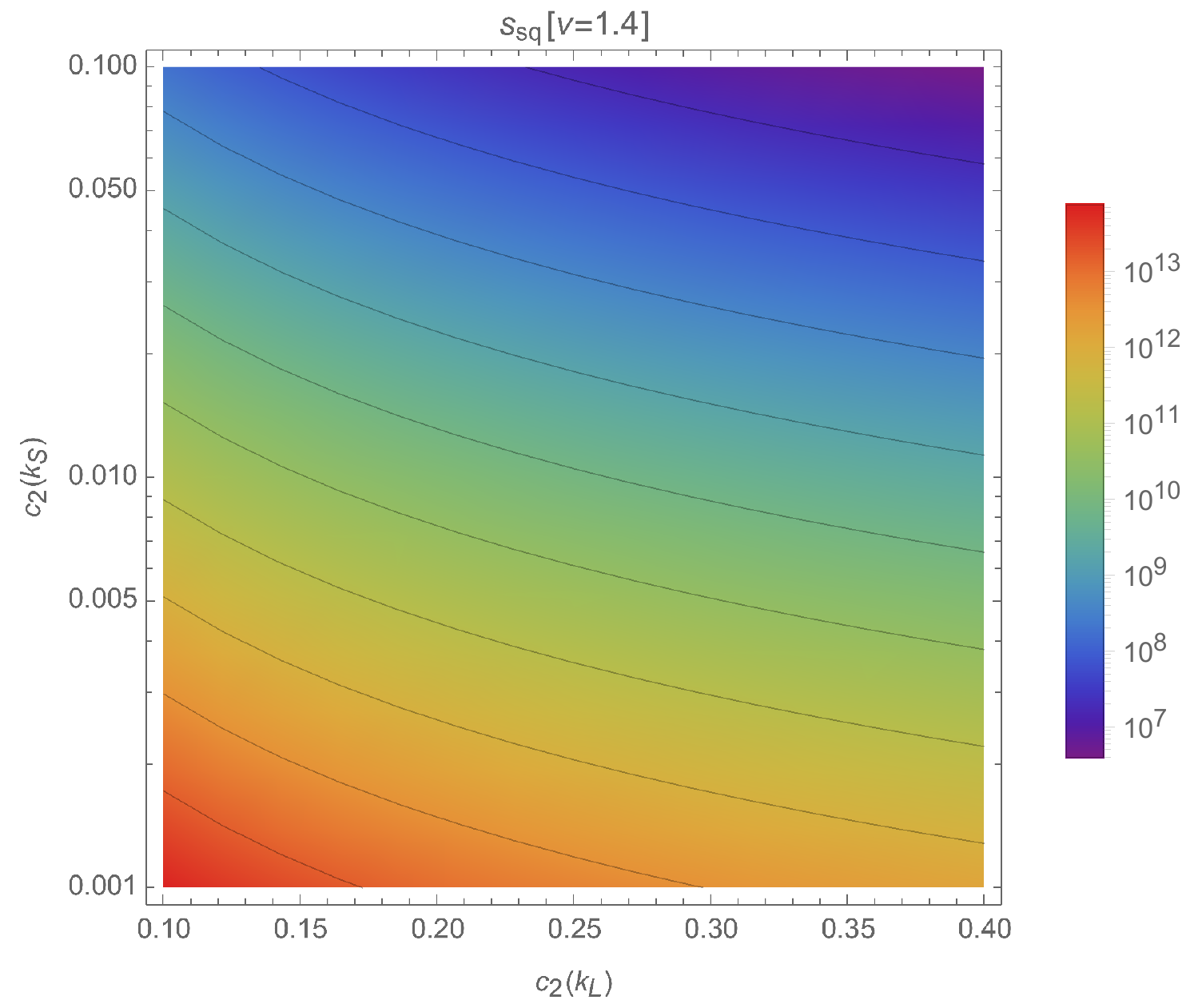}
\caption{Fit of the numerical results obtained for $s_{sq}[\nu=1.4]$ as a function of $c_2(k_S)$ and $c_2(k_L)$, the sound speeds of the short and long scale modes respectively.}
\label{fig:ssq nu 1.4 c2(kL) c2(kS)}
\end{figure}

We now evaluate the bispectrum in the squeezed limit $k_3\ll k_1\sim k_2$ and, for practical purposes, identify $k_3\equiv k_L$ and $k_1\sim k_2\equiv k_S$. We find that the leading contributions to the bispectrum are given by \eqref{FA} and \eqref{FB}, while the other permutations as well as the C term \eqref{FC} are sub-leading. Details on the derivation are included in Appendix \ref{appendix squeezed}. Our findings on tensor non-Gaussianities are somewhat reminiscent of the analysis performed in \cite{Chen:2009zp} for (the scalar sector of) \textit{quasi single field} inflation and in \cite{Dimastrogiovanni:2018gkl} for (the tensor sector of) the EFT set-up. The bispectrum in the squeezed configuration reads
\begin{equation}
\label{Bsigma squeezed}
    B_{(\sigma) sq}(k_L, k_S)= \frac{24\times 2^{\nu} \pi^2}{k_S^{9/2-\nu} k_L^{3/2+\nu}} \frac{\mu}{H} \Big(\frac{\rho}{M_{Pl}} \Big)^3 s_{sq}(\nu,k_L,k_S)\;,
\end{equation}
where 
\begin{equation}
\begin{split}
\label{ssq}
    s_{sq}&(\nu, k_L,k_S)=\frac{\Gamma(\nu)}{c_2(k_L)^\nu}  \int_{-\infty}^{0} dx_1 \;\int_{-\infty}^{x_1} dx_2 \;\int_{-\infty}^{x_2} dx_3  \times \\
 & \Big\{ (-x_2)^{1/2-\nu} (-x_1)^{-1/2}(-x_3)^{-1/2}\; \sin{(-x_1)} \Im\Big[H_\nu^{(1)}(-c_2(k_S) x_1) H_\nu^{(2)}(-c_2(k_S) x_2)\Big] \\
 & \Im\Big[ \mbox{e}^{i x_3 } H_\nu^{(1)}(-c_2(k_S) x_2)  H_\nu^{(2)}(-c_2(k_S) x_3)\Big] +  (-x_1)^{-1/2}(-x_2)^{-1/2} (-x_3)^{1/2-\nu}\\
 &\Im\Big[H_\nu^{(1)}(-c_2(k_S) x_3) H_\nu^{(1)}(-c_2(k_S) x_3) H_\nu^{(2)}(-c_2(k_S) x_1)H_\nu^{(2)}(-c_2(k_S) x_2) \Big]  \sin{(-x_1)} \sin{(-x_2)}\Big\}  \\
 & \times \int_{-\infty}^{0} dy_4 (-y_4)^{-1/2}\Re \Big[\mbox{e}^{-i y_4} H_\nu^{(1)}(-c_2(k_L) y_4) \Big]\;.
\end{split}
\end{equation} 
Similarly to what has been done for the equilateral configuration, the numerical results can be fitted by a power law
\begin{equation}
\label{ansatz power law sq}
 s_{sq} [\nu, c_2(k_L), c_2(k_S)]= \frac{b_\star}{c_2(k_L)^{2\nu} c_2(k_S)^{2\nu} }   \;, 
\end{equation} 
which is used to arrive at Fig.~\ref{fig:ssq nu 1.4 c2(kL) c2(kS)}, where setting $\nu=1.4$ gives
\begin{equation}
\label{fit ansatz nu=1.4}
    s_{sq} [\nu=1.4, c_2(k_L), c_2(k_S)]\simeq  \frac{482.8}{c_2(k_L)^{2.8} c_2(k_S)^{2.8} } \;.
\end{equation}

\begin{figure}
\centering
\captionsetup[subfigure]{justification=centering}
   \begin{subfigure}[b]{0.49\textwidth}
    \includegraphics[width=\textwidth]{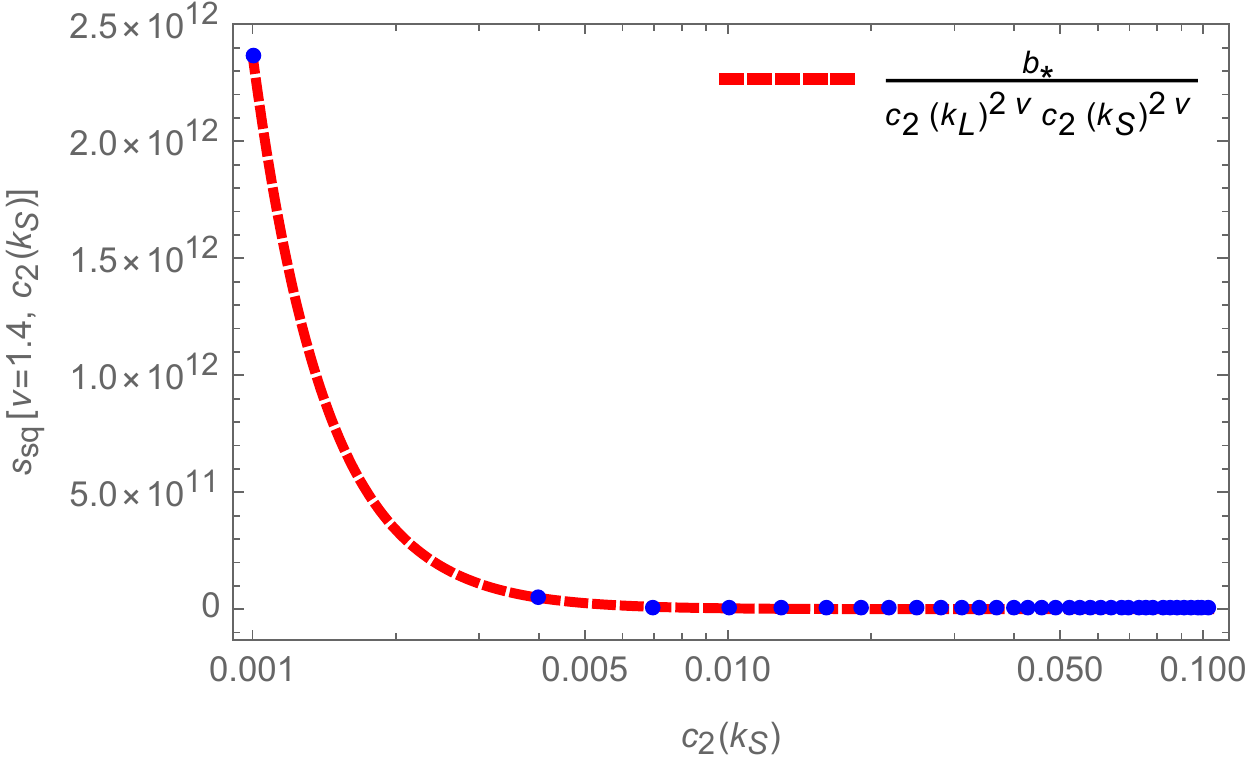}

  \end{subfigure}
  \begin{subfigure}[b]{0.49\textwidth}
    \includegraphics[width=\textwidth]{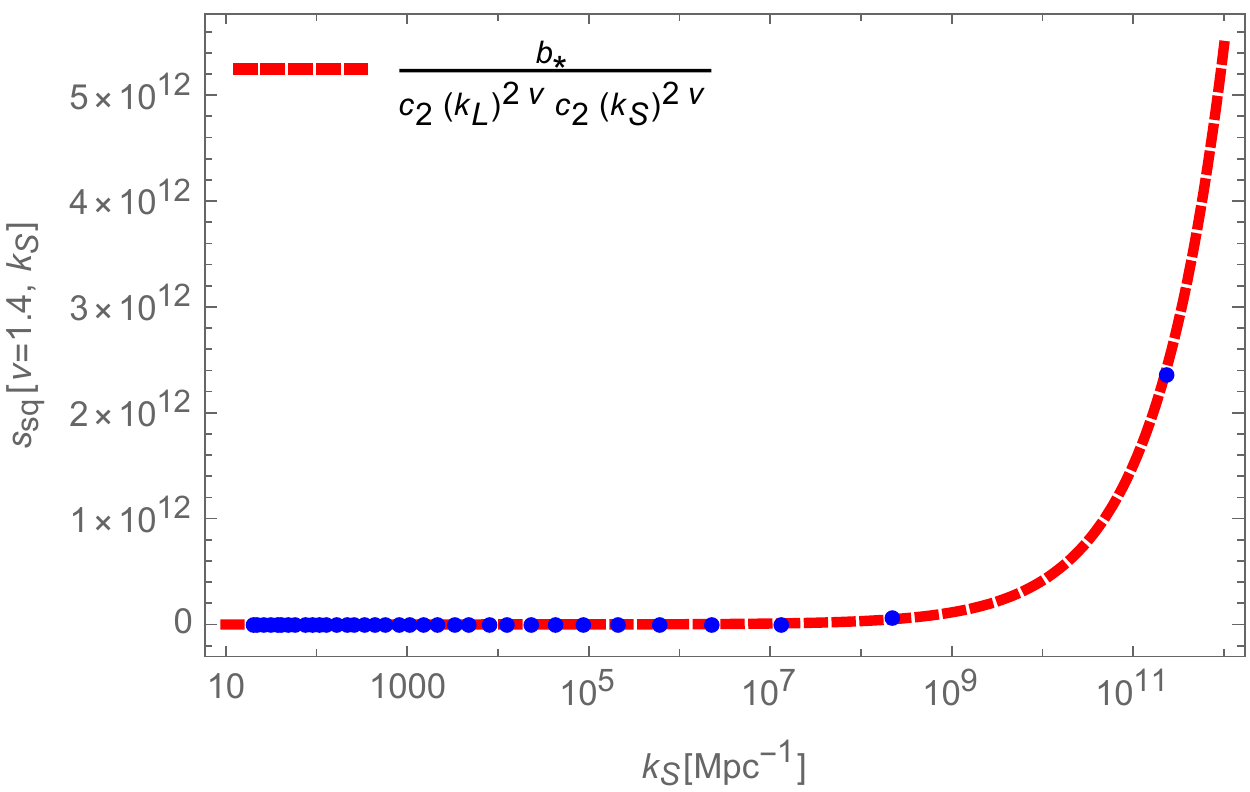}

  \end{subfigure}
  \caption{Results for $s_{sq}(\nu=1.4)$. On the left panel Eq.\eqref{fit ansatz nu=1.4} with $c_2(k_L)=0.346$ is displayed as a function of the value of the sound speed on small scales $c_2(k_S)$, while on the right Eq.\eqref{fit nu=1.4 sq kS} is plotted, with the long mode fixed at CMB scales  and $s_2=-0.2$. In both plots, blue dots represent numerical results.}
  \label{fig:ssq nu 1.4 main}
\end{figure}

\noi In order to visualize our findings in a different fashion, we provide in Fig.~\ref{fig:ssq nu 1.4 main}  the numerical results and the fit \eqref{fit ansatz nu=1.4} with fixed $c_2(k_L)=0.346$. The explicit scale dependence is given by
\begin{equation}
\label{fit nu=1.4 sq kS}
      s_{sq} [\nu=1.4, k_L,\, k_S]\simeq  482.8 \Big(\frac{k_L}{a_0H_0} \Big)^{-2.8 s_2} \Big(\frac{k_S}{a_0H_0} \Big)^{-2.8 s_2} \;, 
\end{equation}
which is plotted on the right in Fig.~\ref{fig:ssq nu 1.4 main} with $k_L\simeq 0.05 \mbox{Mpc}^{-1}$ and $s_2=-0.2$. Just as for the equilateral configuration, a smaller $c_2$ enhances the amplitude of non-Gaussianities. In Appendix \ref{app seq}, a similar analysis is performed for mass values $\nu=\{0.4,\, 0.8, \, 1.1, \, 1.48\}$. The lighter the spin-2 field is ($\nu\rightarrow 3/2$), the greater the amplitude of $s_{sq}(\nu)$.

\subsection{Shape}
\begin{figure}
\centering
\captionsetup[subfigure]{justification=centering}
   \begin{subfigure}[b]{0.49\textwidth}
    \includegraphics[width=\textwidth]{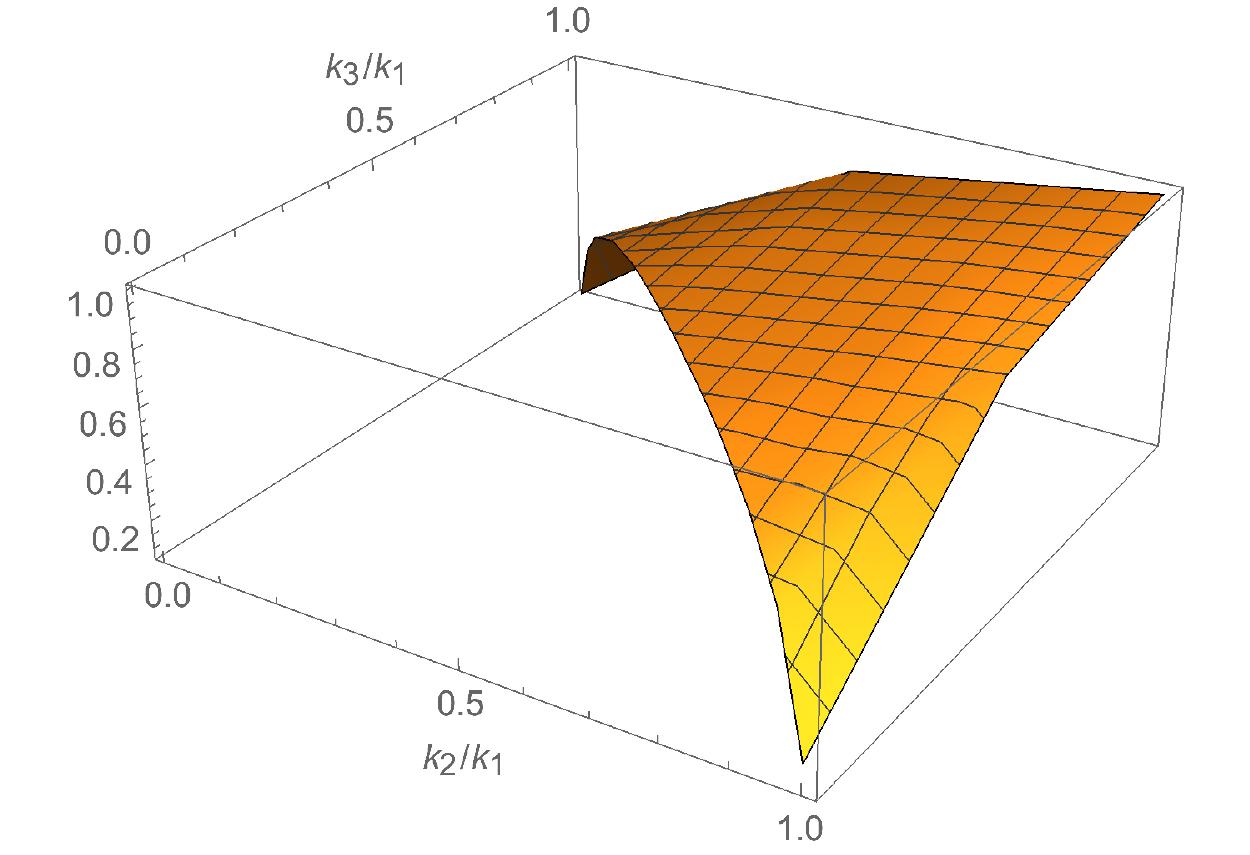}

  \end{subfigure}
  \begin{subfigure}[b]{0.47\textwidth}
    \includegraphics[width=\textwidth]{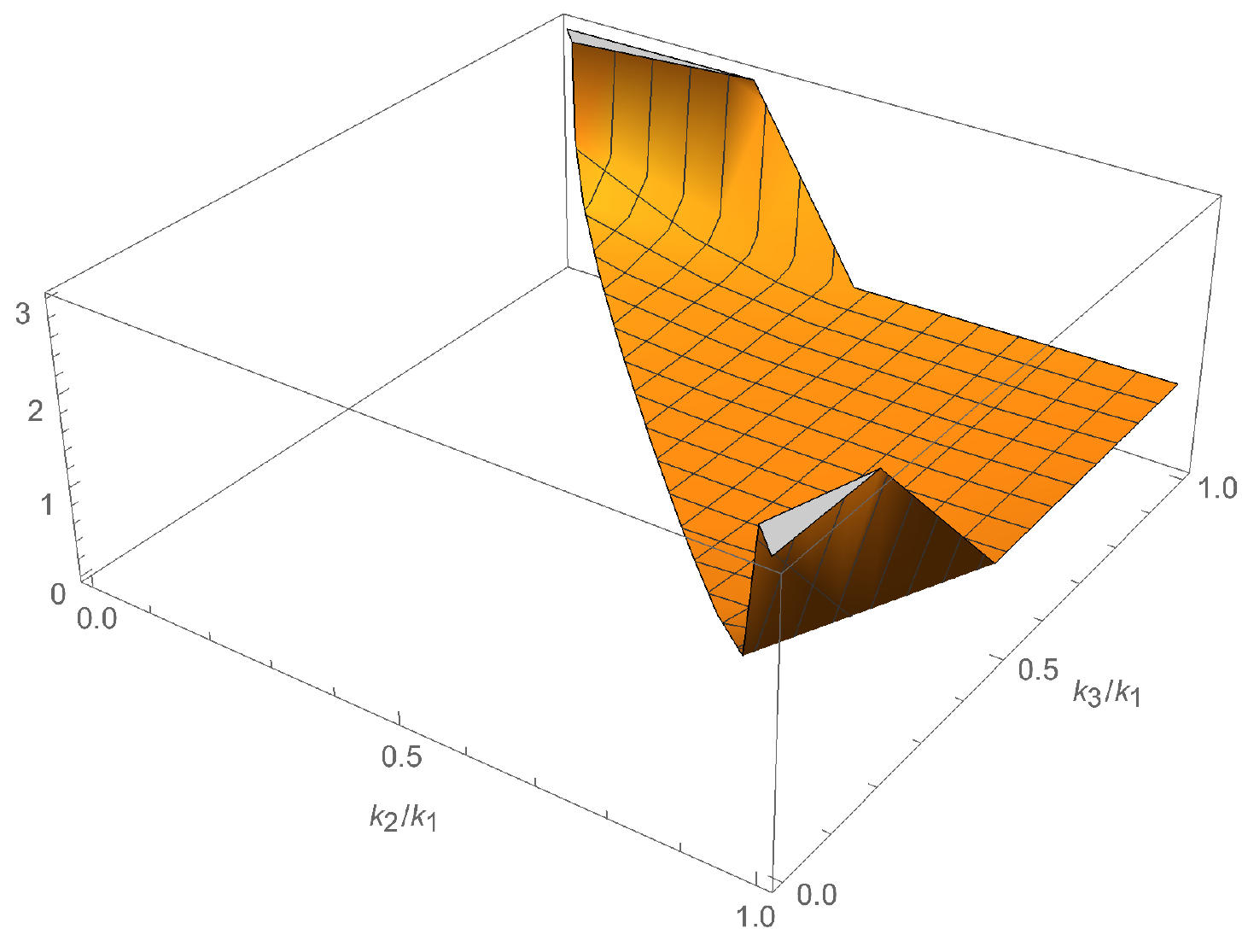}

  \end{subfigure}
  \caption{Shape-function for $\nu=0$ (left) and $\nu=1$ (right). To conform with the literature convention, the bispectrum has been multiplied by $(k_1 k_2 k_3)^2$ and the weight $\mathcal{A}^{\lambda_1 \lambda_2 \lambda_3}$ is not included. The shape values are normalised with respect to the value in the equilateral point $k_1=k_2=k_3$.}
  \label{fig:shapes}
\end{figure}
We move now to study the shape function of the bispectrum, i.e. the dependence on the configuration of the momenta $(k_1,\, k_2,\, k_3)$. We expect it to interpolate between the local and equilateral configurations depending on the mass of the spin-2 
field mediating the interaction in the diagram. This expectation stems from the analogous interactions one finds in the scalar sector of  quasi-single field inflation \cite{Chen:2009zp}. In particular, for a lighter particle, $\nu \gtrsim 1$, the signal peaks in the local\footnote{Strictly speaking, it would be more appropriate to say that the bispectrum peaks in the squeezed limit and that its shape-function is very similar to that obtained by employing the local template. One may define a scalar product between shape functions (see e.g. \cite{Creminelli:2010qf}) and quantify precisely their overlap. It is usually assumed in the literature that an overlap above 75\% would make two templates difficult to distinguish from each other via CMB probes.} configuration, while for smaller value $\nu \ll 1$, i.e. for a heavier field, the bispectrum displays a momentum dependence akin to the equilateral template. As an example, we study the shape-functions for $\nu=0$ and $\nu=1$ in presence of $k$-dependent sound speed $c_2$, with initial condition $c_{2\,in}=1$ and $s_2=-0.2$. These  are plotted in Fig.~\ref{fig:shapes}: on the left for the case $\nu=0$,  and on the right for $\nu=1$. The plots are produced numerically, after applying a Wick rotation to the mixed-form of the bispectrum.

The fact that the shape-function tends towards the equilateral template for interactions mediated by massive particles (as opposed to the light and/or massless fields) has a simple explanation as clear already in the scalar case. The (quasi dS) wave-function for massive fields has approximately a non-zero $(k \tau)^{3/2-\nu}$ factor in front of what would be the massless solution. This term suppresses the wavefunction (and, in turn, the bispectrum) after horizon crossing especially for small wavenumber values, so that the signal in the squeezed configuration is suppressed, to the advantage of the equilateral one. For massless (scalar) fields $\nu=3/2$ so that the same factor is instead equal to unity and therefore inconsequential for the shape. We also note that, despite $c_2$ not being constant in our set-up, the shape-function does not noticeably  change w.r.t. the constant case, unlike the bispectrum amplitude.

\section{Bounds on tensor non-Gaussianities at CMB scales}
\label{sec:CMB bounds}
We now explore the consequences of current bounds on tensor non-Gaussianity, i.e. $f_{\rm NL}^{eq}$ and $f_{\rm NL}^{sq}$ at CMB scales. We shall omit the \textit{tensor} superscript on $f_{\rm NL}$. In particular, the central values and $1\sigma$ error for the equilateral and squeezed template read \cite{Akrami:2019izv, Shiraishi:2019yux}
\begin{equation}
\label{CMB bounds values}
    f_{\rm NL}^{eq}=600 \pm 1600\;,  \;\;\; f_{\rm NL}^{sq}=290 \pm 180 \;.
\end{equation}
We consider the configuration described by the parameters in \eqref{interesting configuration}. As anticipated in Section \ref{sec:theory}, this choice is interesting as it is potentially testable at interferometer scales . The non-linearity parameters in \eqref{CMB bounds values} are defined as 
\begin{gather}
  f_{\rm NL}^{eq} \equiv\frac{B_\gamma^{+++}(k,k,k)}{\frac{18}{5}P_\zeta(k)^2} \\
  f_{\rm NL}^{sq} \equiv \lim_{k_3 \ll k_1 \sim k_2} \frac{B_\gamma^{+++}(k_1,k_2,k_3)}{S^{sq}(k_1,k_2,k_3)}   \;,
\end{gather}
where to connect with the bispectrum definition given in Eq. \eqref{bispectrum def}, we identify $B_\gamma^{+++}(k_1,k_2,k_3) \equiv \mathcal{A}^{RRR}B_{\sigma}(k_1,k_2,k_3)/2\sqrt{2}$. The numerical factor $\mathcal{A}^{RRR}$ is equal to $27/64$ and $1/4$ in the equilateral and squeezed configuration respectively \cite{Dimastrogiovanni:2018gkl}. Note that $f_{\rm NL}^{eq}$ has the same definition as the parameter $f_{\rm NL}^{tens}$ introduced in the Planck team publication \cite{Akrami:2019izv}. In the squeezed limit, the bispectrum shape template $S^{sq}$ reduces to 
\begin{equation}
\label{sq template}
S^{sq} (k_L,\, k_S)=\frac{12}{5} (2\pi^2 \mathcal{P}_\zeta)^2 \frac{1}{k_L^3 k_S^3}\;,
\end{equation}
where $k_L\ll k_S$. The scalar power spectrum is $P_\zeta(k)=\frac{2 \pi^2}{k^3}\mathcal{P}_\zeta(k)$, where $\mathcal{P}_\zeta(k)$ is given in Eq.\eqref{scalar power spectrum}. Equipped with these definitions and by using \eqref{Bsigma equilateral} and \eqref{Bsigma squeezed}, one can calculate the values of $f_{\rm NL}^{eq}$ and $f_{\rm NL}^{sq}$ within the EFT. 
 
\begin{figure}
\centering
\includegraphics[scale=0.7]{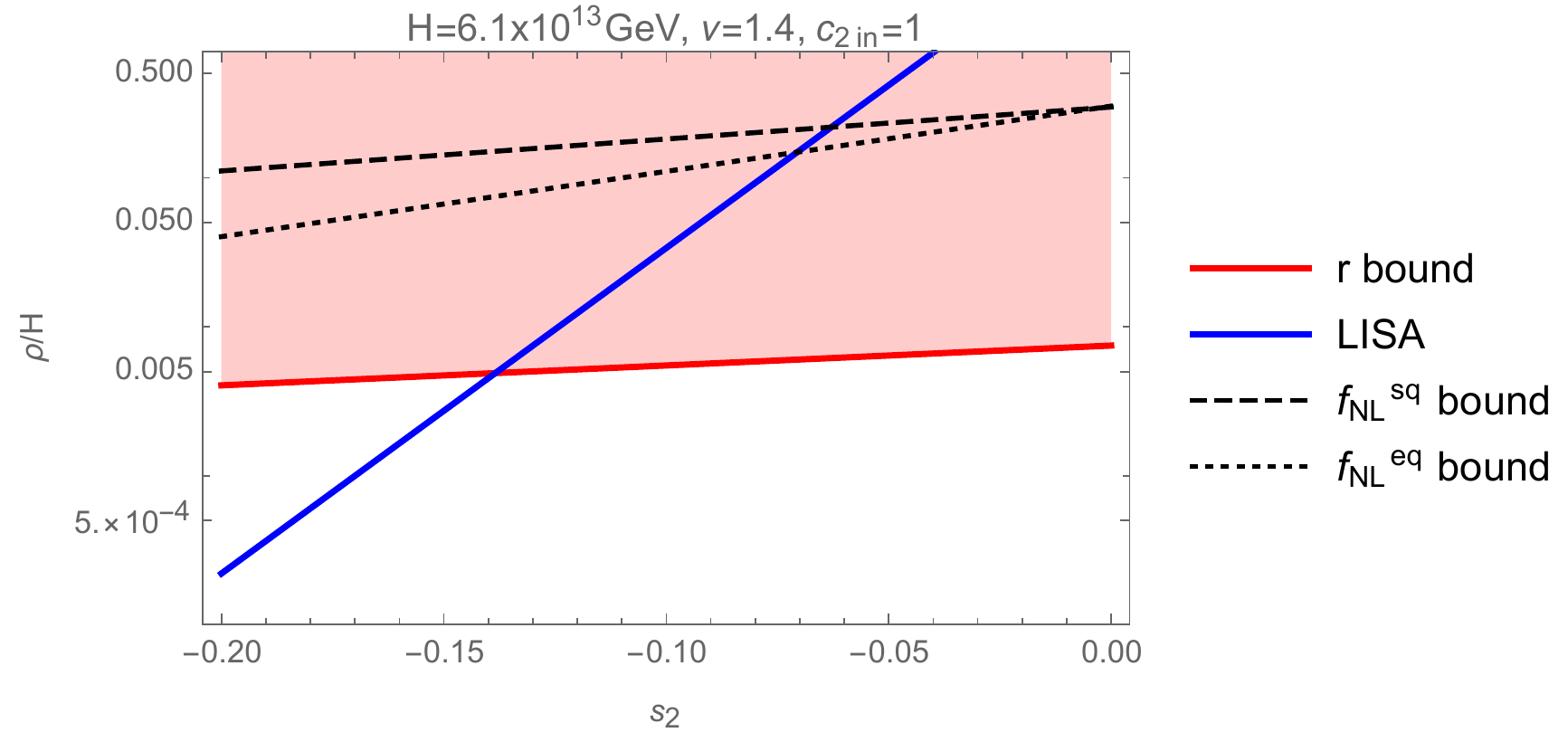}
\caption{Effective theory parameter space $(s_2,\, \rho/H)$ of the configuration $\{ H=6.1\times 10^{13}\,\mbox{GeV}, \, \nu=1.4,\,c_{2\,in}=1 ,\, \mu/H=0.5 \}$. Bounds in \eqref{CMB bounds values} are plotted with dashed lines. Those lines lie in the red-shaded region, which is excluded already by the bound on the tensor-to-scalar ratio $r$. The area above the blue line will be surveyed by LISA. For more details on the construction of the parameter space see \cite{Iacconi:2019vgc}.}
\label{fig:par space with CMB bounds}
\end{figure}
\noi In Fig.~\ref{fig:par space with CMB bounds}, the bounds at large scales  \eqref{CMB bounds values} are displayed on the parameter space $(s_2,\, \rho/H)$ of the configuration \eqref{interesting configuration}. The additional blue and red lines in the plot represent the strongest existing bound, which comes from the limit on the tensor-to-scalar ratio at CMB scales ($r<0.056$)  \cite{Akrami:2018odb}, and the line corresponding to LISA sensitivity: the area above the blue line is surveyable by LISA. The bounds from Eq.~\eqref{CMB bounds values} are weaker on the parameter space than the constraint coming from the current upper limit on $r$. 
\begin{figure}
\centering
\captionsetup[subfigure]{justification=centering}
   \begin{subfigure}[b]{0.49\textwidth}
    \includegraphics[width=\textwidth]{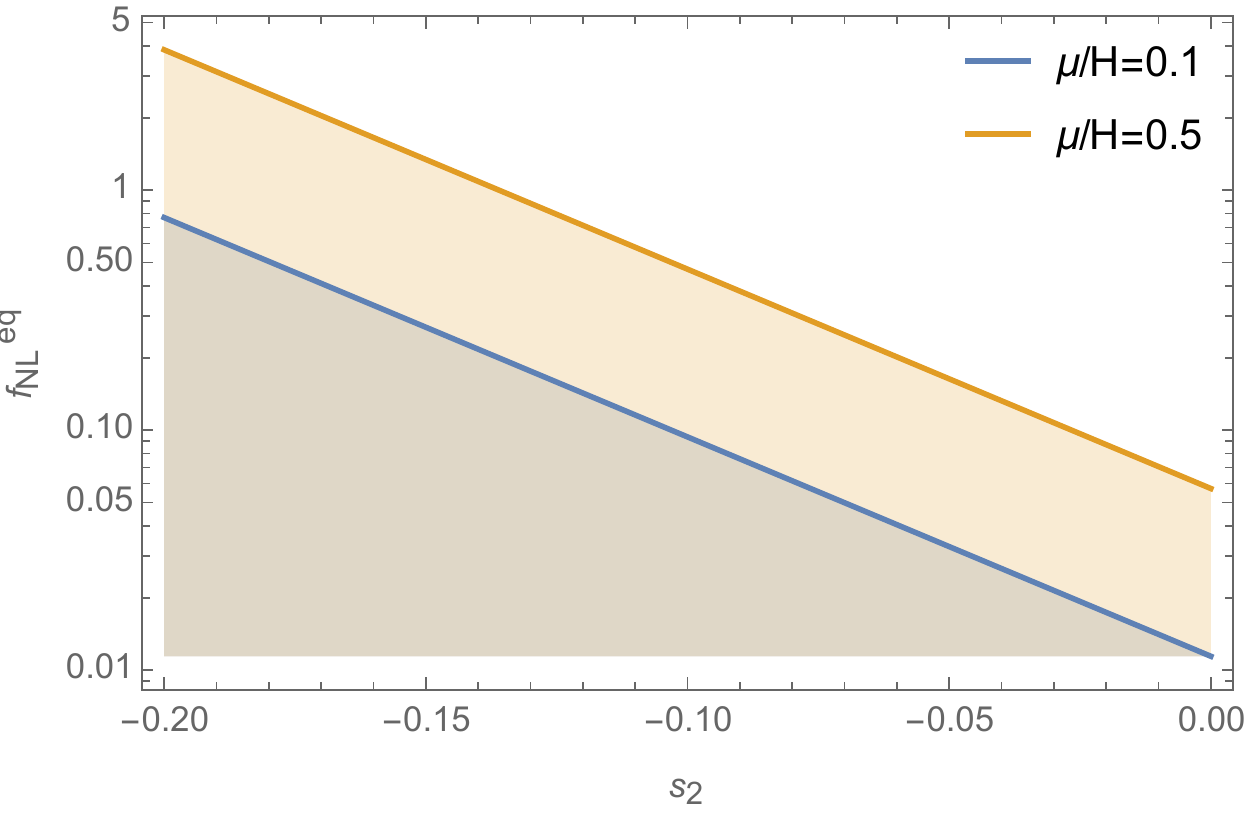}

  \end{subfigure}
  \begin{subfigure}[b]{0.49\textwidth}
    \includegraphics[width=\textwidth]{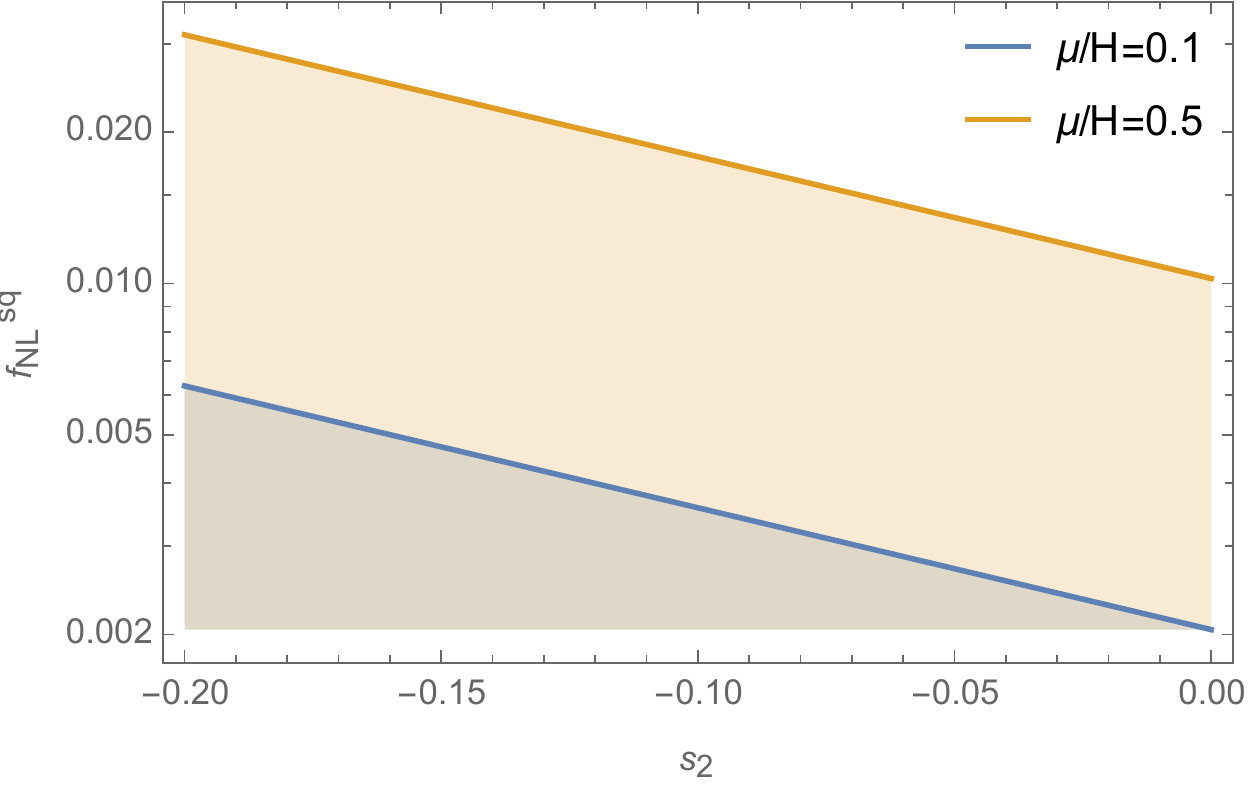}

  \end{subfigure}
  \caption{Maximum level of tensor non-Gaussianities produced at $k_{\rm CMB}=0.05\, \mbox{Mpc}^{-1}$ in the set-up $ \{H=6.1\times 10^{13}\mbox{GeV}, \, \nu=1.4, \, c_{2\, in}=1\}$. $f_{\rm NL}^{eq}$ and $f_{\rm NL}^{sq}$ are represented on the left and right panels respectively, for different values of the cubic self-interaction coupling $\mu/H$. }
  \label{fig:fNLmax at CMB scales}
\end{figure}

Given an upper bound on $\rho/H$ as a function of $s_2$ obtained by requiring $r<0.056$, it is possible to maximize the level of tensor non-Gaussianities produced at CMB scales for the configuration under scrutiny. The corresponding amplitudes $f_{\rm NL}^{eq}$ and $f_{\rm NL}^{sq}$ are given in Fig.~\ref{fig:fNLmax at CMB scales}. The behavior with respect to $s_2$ is clear: the greater $|s_2|$ is, the faster $c_2$ decreases (see Fig.~\ref{fig:c2(k)eq}) and a smaller sound speed enhances the level of non-Gaussianities, as shown in Section \ref{sec:calculation}. Although we have to conclude that the present bound on $r$ is more constraining on the region of parameter space we are probing than bounds on non-Gaussianity, one should not infer that this holds for the entire parameter space of the EFT. Our findings are specific to our starting points in terms of the chosen parameters as well as the (negligible by choice) role played by the helicity-0 mode in sourcing the scalar signal. Our choice of the parameter space region to inspect has been guided by its testability at small scales by upcoming probes, and is by no means representative of the full  EFT Lagrangian phenomenology.

\section{Testing squeezed GWs non-Gaussianity at small scales}
\label{sec:detectability}
As shown in the last Section, tensor non-Gaussianities produced within the configuration in Eq.~\eqref{interesting configuration} are well-below current bounds at CMB scales. When it comes to testing inflationary GW higher-point correlators at small scales, one should be aware that these are not directly testable: de-correlation sets in as a result of the propagation through structure that GWs undergo on their way to the detector \cite{Bartolo:2018evs}.

Nonetheless, it is possible to test non-Gaussianities in a specific configuration, namely the \textit{ultra-squeezed} one. Such nomenclature refers to the case where the long wavenumber is (nearly) horizon size or larger, so that it avoids propagation effects whilst still correlating with short, well-inside-the-horizon, modes. The specific effect of long tensor fluctuation is to induce, in the presence of non-trivial\footnote{Here ``non-trivial'' does not mean merely non-zero. The squeezed limit of the three-point function is directly physical whenever so-called consistency relations (CRs) are broken \cite{Hinterbichler:2013dpa} , i.e. whenever the squeezed three-point function cannot be expressed as the action of a gauge transformation on the corresponding power spectrum. The prototypical case of broken CRs is that of multi-field inflation. However, a multi-field scenario does not by itself guarantee CRs breaking. A quick route to see that CRs are indeed broken in our set-up when the bispectrum contribution is mediated by $\sigma$ is to notice that such interactions are regulated by the parameter $\mu$ (see Eq.~\eqref{L3}), which does not appear in the quadratic Lagrangian. The reader familiar with quasi-single field inflation may take another path to the same conclusions by noticing the similarities between the quantity $\mu$ here and (the third derivative of) the potential $V(\sigma)$ of the extra  field $\sigma$ in \cite{Chen:2009zp}.} ultra-squeezed tensor non-Gaussianity, a quadrupolar anisotropy on the power spectrum of the short modes \cite{Jeong:2012df,Dai:2013kra,Brahma:2013rua,Dimastrogiovanni:2014ina,Dimastrogiovanni:2015pla}. This idea has been explored in the context of inflationary GW at small scales in \cite{Ricciardone:2017kre,Dimastrogiovanni:2019bfl,Fujita:2019tov}. One should also keep in mind that, next to the cosmological SGWB we want to probe, there is an astrophysics SGWB whose signal we need to disentangle from the primordial one. For a comprehensive account on how to characterise the anisotropies of the stochastic GWs background, we refer the interested reader to recent work on the topic \cite{Contaldi:2016koz,Bartolo:2019oiq,Bartolo:2019yeu}. It suffices here to say that a sufficiently large primordial signal at small scales may dominate the anisotropic component \cite{Dimastrogiovanni:2019bfl}.

In what follows we briefly review the results of \cite{Dimastrogiovanni:2019bfl} and then explore their consequences for the EFT set-up at hand. This is appropriate given that the EFT bispectrum has a significant squeezed component for sufficiently light $\sigma$, such as is the case for e.g. $\nu=1$ and $\nu=1.4$. In the presence of a non-trivial ultra-squeezed primordial tensor bispectrum, a long tensor mode $k_L$ induces on the tensor power spectrum  evaluated locally at $\mathbf{x_c}$ a quadrupolar modulation of the form

\begin{equation}
    P_\gamma(\mathbf{k_S}, \mathbf{x_c})|_{k_L}=  P_\gamma(k_S) \Big(1+ \mathcal{Q}_{lm}(\mathbf{k_S}, \mathbf{x_c})\hat{k}_{S\,l} \hat{k}_{S\,m} \Big) \;,
\end{equation}
where $P_\gamma(k)$ is the standard isotropic component of the power spectrum, $k_S$ stands for a generic small wavelength such that $k_S\gg k_L$, and $\mathcal{Q}_{lm}$ is the anisotropy parameter defined as \begin{equation}
    \mathcal{Q}_{lm}(\mathbf{k_S}, \mathbf{x_c})\equiv \int \frac{d^3k_L}{(2\pi)^3} \, \mbox{e}^{i \,\mathbf{k_L} \cdot \mathbf{x_c}}\, F_{\rm NL}(k_L, k_S) \sum_{\lambda_3} \epsilon_{lm}^{\lambda_3} (-\hat{k}_L)\gamma_{-\mathbf{k_L}}^{*\, \lambda_3} \;.
\end{equation}
The quantity $  F_{\rm NL}(k_L, k_S) $ is the non-Gaussianity parameter in the squeezed configuration, defined as 
\begin{equation}
\label{fNL(k_L,k_S) def}
    F_{\rm NL}(k_L, k_S) \equiv  \frac{B_{sq}(k_L, k_S)}{P_\gamma(k_L) P_\gamma(k_S) } \;,
\end{equation}
{where $P_\gamma(k)=2\pi^2 \mathcal{P}_\gamma(k) /k^3$ and the quantities $\mathcal{P}_\gamma(k)$ and $B_{sq}(k_L, k_S)$ are spelled out in Eqs.\eqref{tensor power spectrum} and \eqref{Bsigma squeezed} respectively.} One can characterize the quadrupolar tensor anisotropy by computing its variance \cite{Jeong:2012df}:
\begin{equation}
\label{variance def}
    \bar{\mathcal{Q}^2} \equiv \langle \sum_{m=-2}^{+2} |\mathcal{Q}_{2m}|^2\rangle  = \frac{8 \pi}{15} \langle \mathcal{Q}_{ij} \mathcal{Q}^{*\,ij}\rangle      \;,
\end{equation}
with
\begin{equation}
\label{QQ}
   \langle \mathcal{Q}_{ij} \mathcal{Q}^{*\,ij}\rangle =16 \int \frac{d^2 \hat{k}_L}{4\pi} \int_{k_{L}^{\rm min}}^{k_{L}^{\rm max}} \frac{dk_L}{k_L} \, F_{\rm NL}^2(k_L, k_S)\, \mathcal{P}_\gamma(k_L) \;,
\end{equation}
where $\mathcal{P}_\gamma(k)$ is the dimensionless tensor power spectrum.  We now use the results in Section \ref{sec:calculation}, configuration~\eqref{interesting configuration}, to explore small-scale signatures associated to the presence of an extra\footnote{``Extra'' with respect to the standard massless spin-2 particle, the graviton, of general relativity.} spin-2 field during inflation. We compute  $\sqrt{\bar{\mathcal{Q}^2}}$ and identify in the EFT parameter space areas that (i)  support a detectable tensor power spectrum and (ii) whose squeezed tensor bispectrum produces a quadrupolar modulation with $\sqrt{\bar{\mathcal{Q}^2}}\gtrsim 0.01$. We use the percent value for anisotropies as a benchmark point. There is ongoing research focussed on establishing whether this will be attainable with upcoming probes (see \cite{Ozsoy:2019slf} and references therein). We should stress at this stage that,  although our analysis has been mainly motivated by the possibility to explore the capability of laser interferometers  to detect inflationary signatures, our results apply equally well to pulsar timing arrays.

\medskip
\begin{figure}
\centering
\captionsetup[subfigure]{justification=centering}
   \begin{subfigure}[b]{0.47\textwidth}
    \includegraphics[width=\textwidth]{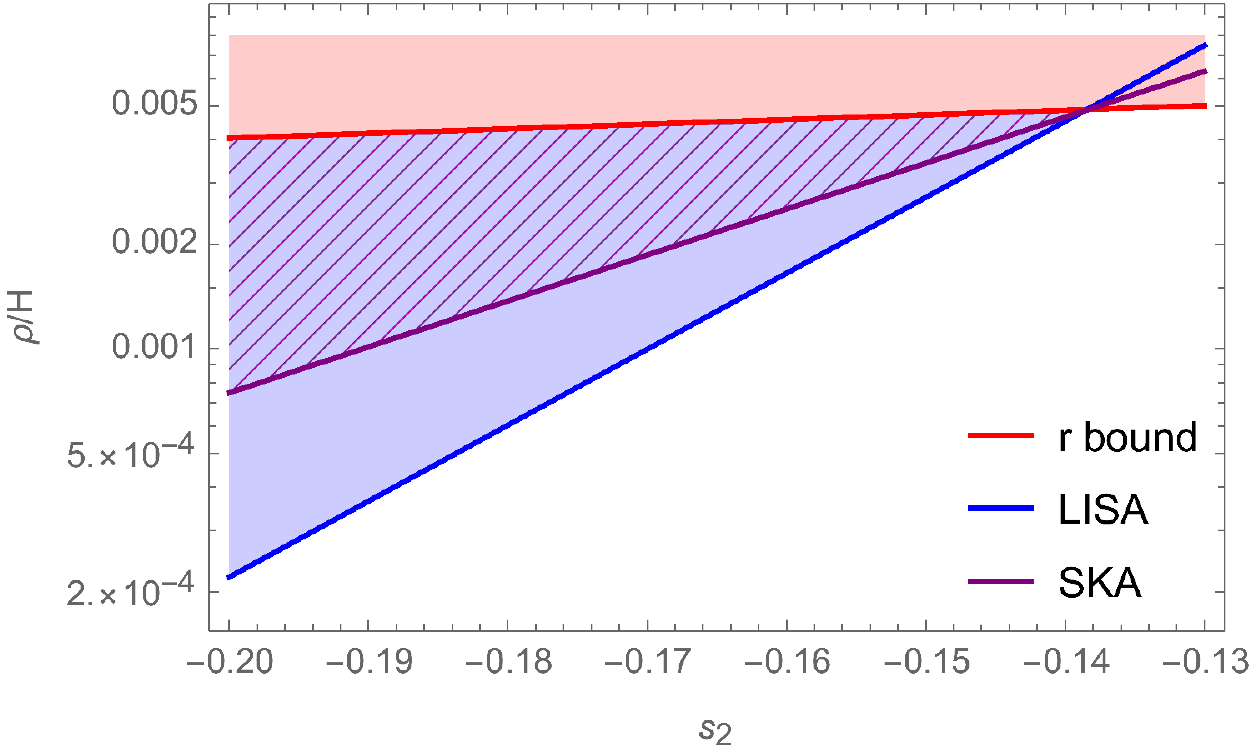}

  \end{subfigure}
  \begin{subfigure}[b]{0.45\textwidth}
    \includegraphics[width=\textwidth]{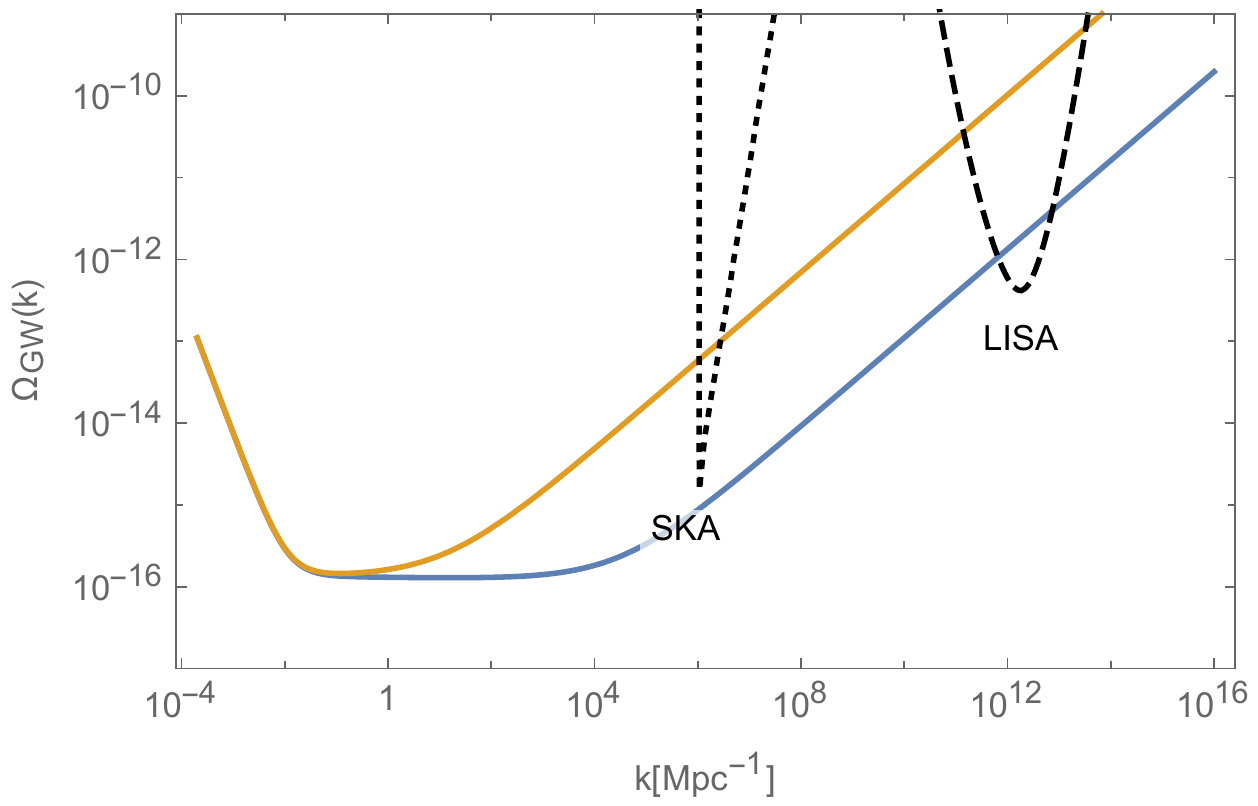}

  \end{subfigure}
  \caption{\textbf{Left panel:} Effective Theory parameter space $(s_2,\, \rho/H)$ of the configuration $\{H=6.1\times 10^{13}\,\mbox{GeV},\, \nu=1.4,\, c_{2\,in}=1\}$. The red-shaded area is excluded by the bound on the tensor-to-scalar ratio $r$. The blue area is surveyable by LISA, on top of which the region highlighted with purple hatch shading is also visible to SKA. \textbf{Right panel:} Examples of two tensor signals sourced within the theory. The orange line corresponds to $(s_2=-0.2,\, \rho/H=0.0035)$ and is visible both to LISA and SKA, while the blue line corresponds to $(s_2=-0.2,\, \rho/H=0.0004)$ and might be detected by LISA only.}
  \label{fig:LISA SKA signals}
\end{figure}
In Fig.~\ref{fig:LISA SKA signals} we plot, on the left side, a specific section of the EFT parameter space: the plane $(\rho/H, s_2)$. Highlighted in blue is the area delivering a GW signal testable by LISA. The area above the purple line in instead at reach for SKA \cite{Janssen:2014dka}. The region above the red line is off-limits as it correspond to a tensor to scalar ratio already excluded by CMB data. The right side of Fig.~\ref{fig:LISA SKA signals} illustrates how two points in parameter space engender a GW signal that is sufficiently large for (i) detection by SKA and LISA or (ii) detection by LISA only. In order to generate the plot, we have used $k_{\rm SKA}=6.5 \times 10^{5}\,\mbox{Mpc}^{-1}$  and $k_{\rm LISA}=10^{12}\,\mbox{Mpc}^{-1}$. For studies on reconstructing the tensor power spectrum with LISA and PTA see \cite{Caprini:2019pxz} and \cite{Tsuneto_2019} respectively. In order to arrive at Fig.~\ref{fig:LISA SKA signals}, we employed the following expression for the GW energy density today \begin{equation}
    \Omega_{GW}(k)=\frac{1}{12}\,\Big(\frac{k}{a_0H_0}\Big)^2\, \mathcal{P}_\gamma(k) \, T^2(k) \;,
\end{equation}
where  $T(k)$ is the standard transfer function.  
\begin{figure}
\centering
\includegraphics[scale=0.7]{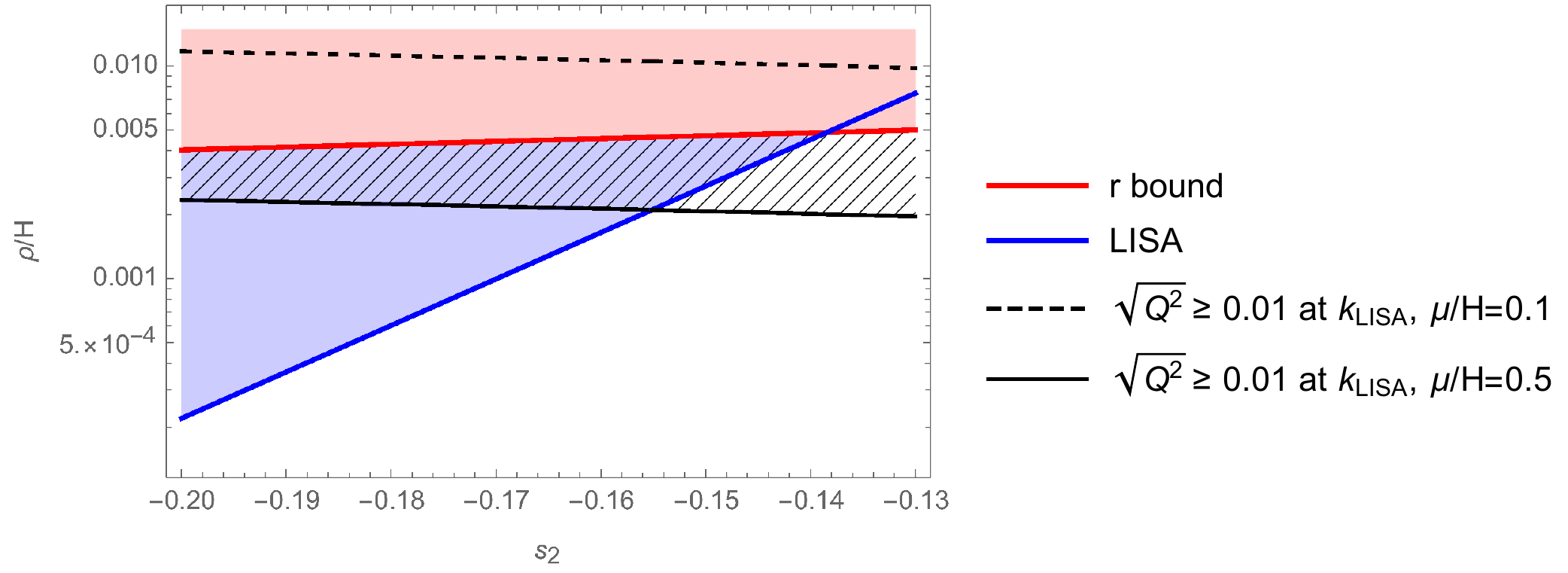}
\caption{Effective Theory parameter space $(s_2,\, \rho/H)$ of the configuration $\{H=6.1\times 10^{13}\,\mbox{GeV},\, \nu=1.4,\, c_{2\,in}=1\}$. The blue area delivers a tensor power spectrum detectable by LISA. The hatch shaded region above the black line corresponds to parameter values which produce a quadrupolar modulation of the tensor power spectrum with standard deviation $\sqrt{\bar{\mathcal{Q}^2}}\geq 0.01$ at LISA scales, with $\mu/H=0.5$. Therefore, if LISA will be able to detect quadrupolar modulations with standard deviation $\geq 0.01$, the squeezed bispectrum can be indirectly tested in the parameter space area which is both hatch and blue shaded. On the other hand, the parameter choice $\mu/H=0.1$ lies in a region which is already excluded by the bound on the tensor-to-scalar ratio.}
\label{fig:par space LISA with Q}
\end{figure} 
Let us now turn to identifying the area of the parameter space delivering a tensor quadrupolar anisotropy with standard deviation of the order of a few percent.

 We focus on LISA first. Using Eqs. \eqref{tensor power spectrum}, \eqref{Bsigma squeezed} and \eqref{fNL(k_L,k_S) def}  in Eq.\eqref{QQ}, one arrives at the value of $\sqrt{\bar{\mathcal{Q}^2}}$. In Fig.~\ref{fig:par space LISA with Q}, the area above the black lines produces a signal with $\sqrt{\bar{\mathcal{Q}^2}}\geq 0.01$; the continuous and dashed lines correspond to, respectively, $\mu/H=0.5$ and $\mu/H=0.1$. The overlap with the  blue area selects the parameter values in the $(s_2,\,\rho/H)$ plane that deliver a detectable tensor power spectrum with a quadrupolar modulation characterised by $\sqrt{\bar{\mathcal{Q}^2}}\geq 0.01$. Depending on the configuration parameters, the $F_{\rm NL}$ values needed to produce a quadrupolar modulation at the percent level are of order $10^3-10^4$. This goes to show how probes such as LISA will, by testing anisotropies, access information on (the size of) squeezed tensor non-Gaussianities and, in turn, the inflationary particle content.
\\We have stressed throughout this paper that the EFT of inflation framework is ideal for capturing the full spectrum of possible signatures of inflationary models. On the other hand, it may be difficult, once a specific observational feature has been identified, to map it back all the way to a precise model of inflation. Indeed, the EFT enables one to associate signatures with specific operators in the Lagrangian of the effective theory of fluctuations around an FRW solution, but it is less illuminating in identifying the complete theory (both background and fluctuations) supporting the acceleration mechanism.  
These considerations apply to the use of the EFT of inflation both in the single-field as well as in the multi-field context. In the latter case however, especially as particles of increasing spin are considered, it is sometimes difficult to arrive at a fully non-linear Lagrangian formulation of the theory (this is the case for higher spin fields). We should nevertheless be aware of the crucial extra step necessary to build a clear-cut signature-to-theory dictionary.

\begin{figure}[h!]
\centering
\includegraphics[scale=0.7]{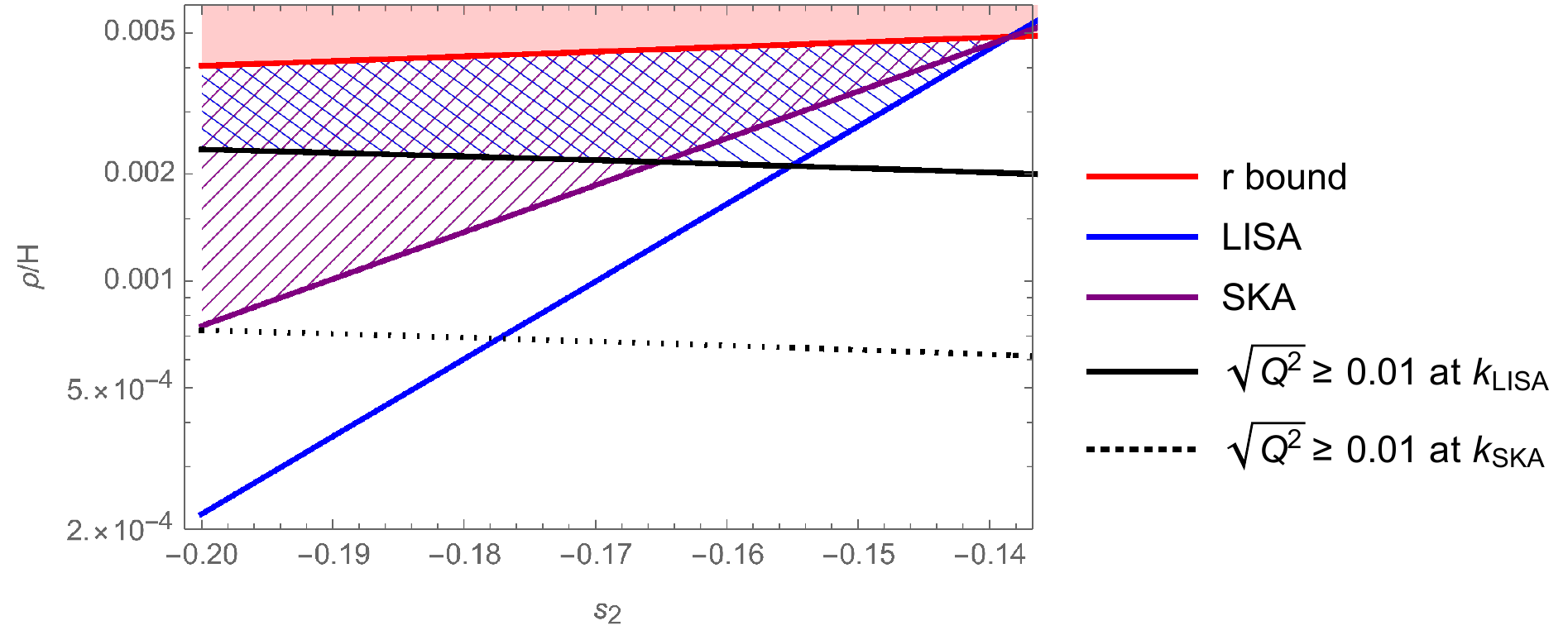}
\caption{Effective Theory parameter space $(s_2,\, \rho/H)$ of the configuration $\{H=6.1\times 10^{13}\,\mbox{GeV},\, \nu=1.4,\, c_{2\,in}=1, \, \mu/H=0.5\}$. The hatch shaded areas deliver a tensor power spectrum detectable by the corresponding probe, with a quadrupolar modulation induced by squeezed tensor non-Gaussianities with standard deviation $\geq 0.01$. The purple and blue colors correspond to SKA and LISA respectively.}
\label{fig:par space LISA SKA with Q}
\end{figure}
In Fig.~\ref{fig:par space LISA SKA with Q}, a similar analysis to the one done for LISA is performed for SKA. The area marked by both blue and purple lines delivers a tensor power spectrum detectable by LISA and SKA with a quadrupolar modulation such that $\sqrt{\bar{\mathcal{Q}^2}}\geq 0.01$.  It is important to point out \footnote{We are grateful to Gianmassimo Tasinato for underscoring the importance of these limitations and for pointing us to the relevant literature.} at this stage the following fact: very recent work \cite{Alonso:2020rar} suggests that, in order to be able to detect anisotropies, the monopole signal should be above the instrument (e.g. LISA) sensitivity curve of about one order of magnitude. A similar analysis exists also for PTAs \cite{Mingarelli:2013dsa}. While the parameter space on the left half of the plot in Fig. \ref{fig:par space LISA SKA with Q} can satisfy this condition, this is not the case towards smaller values of $|s_2|$. Our analysis underscores the  possibility of testing the same signal with different probes and on different scales. The multi-probe characterisation of the GW signal is a crucial steps towards solving the cosmological vs astrophysical sources dichotomy.

\section{Conclusions}
\label{sec:conclusions}
The quest for a deeper understanding of inflationary dynamics is certainly worthwhile pursuit in its own right: in doing so we are, after all, probing the origin of the universe. The current status of cosmology and related fields makes it, if possible, even more timely and appealing. A growing number of experimental missions will search for imprints of primordial physics across an unprecedented range of scales. Their ever-improving sensitivities attest to the fact that this is indeed the era of precision cosmology. The potential for progress in early universe physics to also impact particle physics is immense: with an energy scale that can be many orders of magnitude above those reached in particle colliders, inflation is a precious portal into Beyond the Standard Model physics. 

In this work we studied the signature of an inflationary scenario equipped with a particle content that goes beyond that of the minimal single-field slow-roll paradigm. By employing an effective field theory approach, we accounted for an extra spin-2 field non-minimally coupled to the inflaton. Such direct couplings weaken what would otherwise be very stringent bounds on the allowed spin-2 mass range, and open up possible signatures in cosmological correlators. The focus of our analysis has been on gauging the capability of small-scale probes of gravity, such as SKA and LISA, to uncover signatures of inflationary dynamics in the gravitational waves spectrum we may observe today.

After reviewing how the EFT parameter space supports a detectable GW signal at small scales once we allow time-dependence for the sound speed of helicity-2 fluctuations, we studied the tensor three-point function. Its amplitude and, most importantly, its shape dependence contain tell-tale signs of the mass (and the couplings) of the extra spin-2 field. We singled out the configurations corresponding to a non-trivial squeezed bispectrum and showed also how this may be indirectly tested at small scales by the anisotropies induced in the GW power spectrum. We quantified the amount of tensor non-Gaussianity needed for it to generate a percent level anisotropy in the GW signal within reach of SKA and LISA.

It will be interesting to also study  squeezed scalar-tensor-tensor non-Gaussianities within the EFT framework. Indeed, as recently shown in \cite{Adshead:2020bji}, the correlation of CMB temperature anisotropies with the stochastic GWs background (anisotropies) on small scales provides a new path to testing the inflationary particle zoo and, crucially, distinguishing the primordial SGWB from the astrophysical one. Naturally, the EFT formalism we have been employing is ideal to extend the analysis to different and additional particle content, including higher-spin fields. We leave this to future work.

\acknowledgments
We are delighted to thank Ema Dimastrogiovanni for collaboration in the early stage of this work and for many illuminating conversations. We are also grateful to Gianmassimo Tasinato for insightful conversations and comments. HA, MF, LI, and DW are supported in part by STFC grants ST/S000550/1 and ST/R505018/1. 

\begin{appendices}

\section{Results for the $s_{eq}(\nu)$ and $s_{sq}(\nu)$ computations}
\label{app seq}
\begin{figure}[!ht]
\centering
\captionsetup[subfigure]{justification=centering}
  \begin{subfigure}[b]{0.40\textwidth}
    \includegraphics[width=\textwidth]{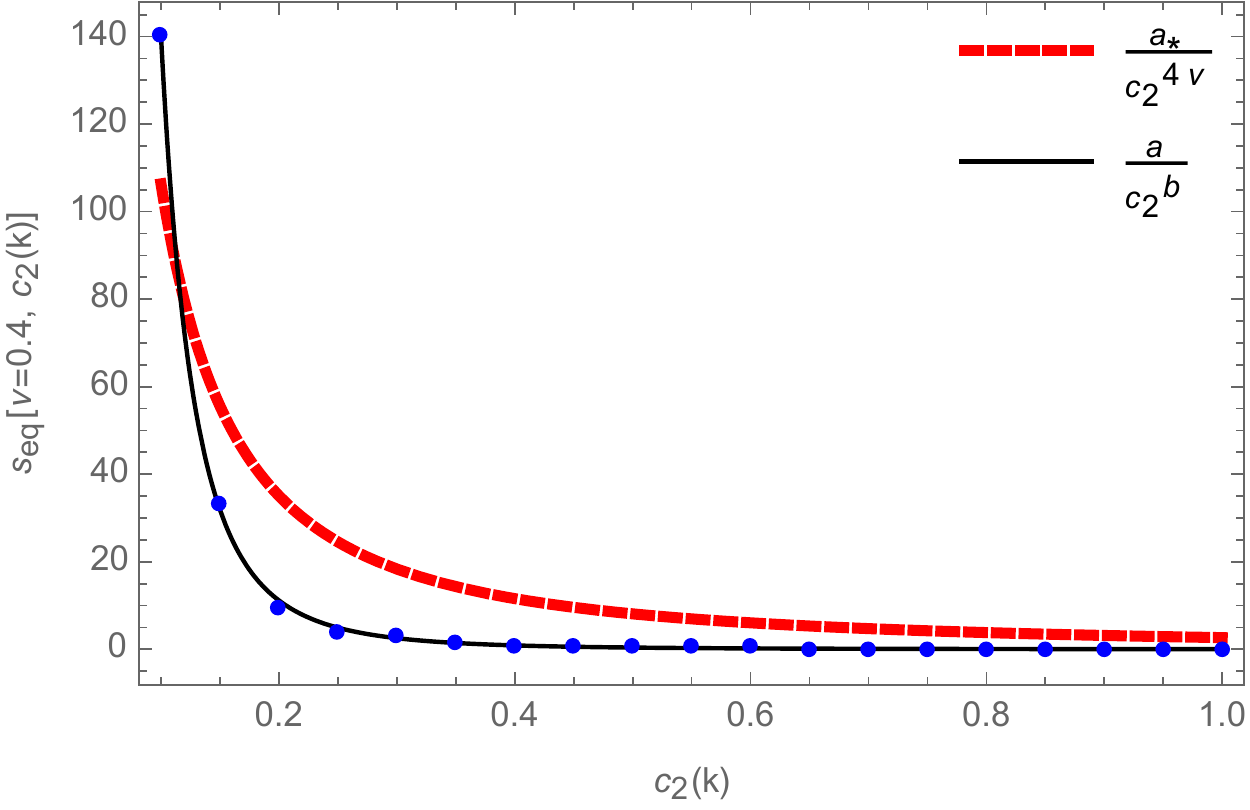}
  \end{subfigure}
  \begin{subfigure}[b]{0.40\textwidth}
    \includegraphics[width=\textwidth]{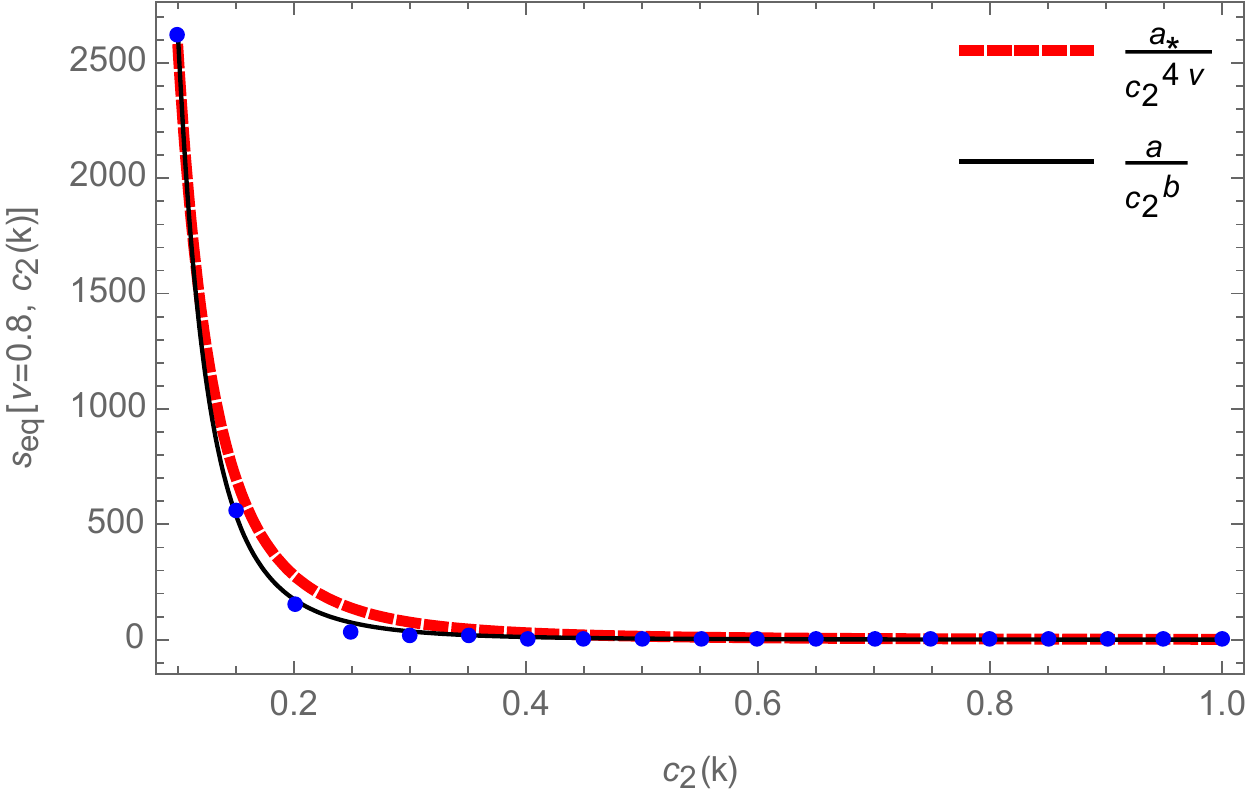}
  \end{subfigure}
  \begin{subfigure}[b]{0.40\textwidth}
    \includegraphics[width=\textwidth]{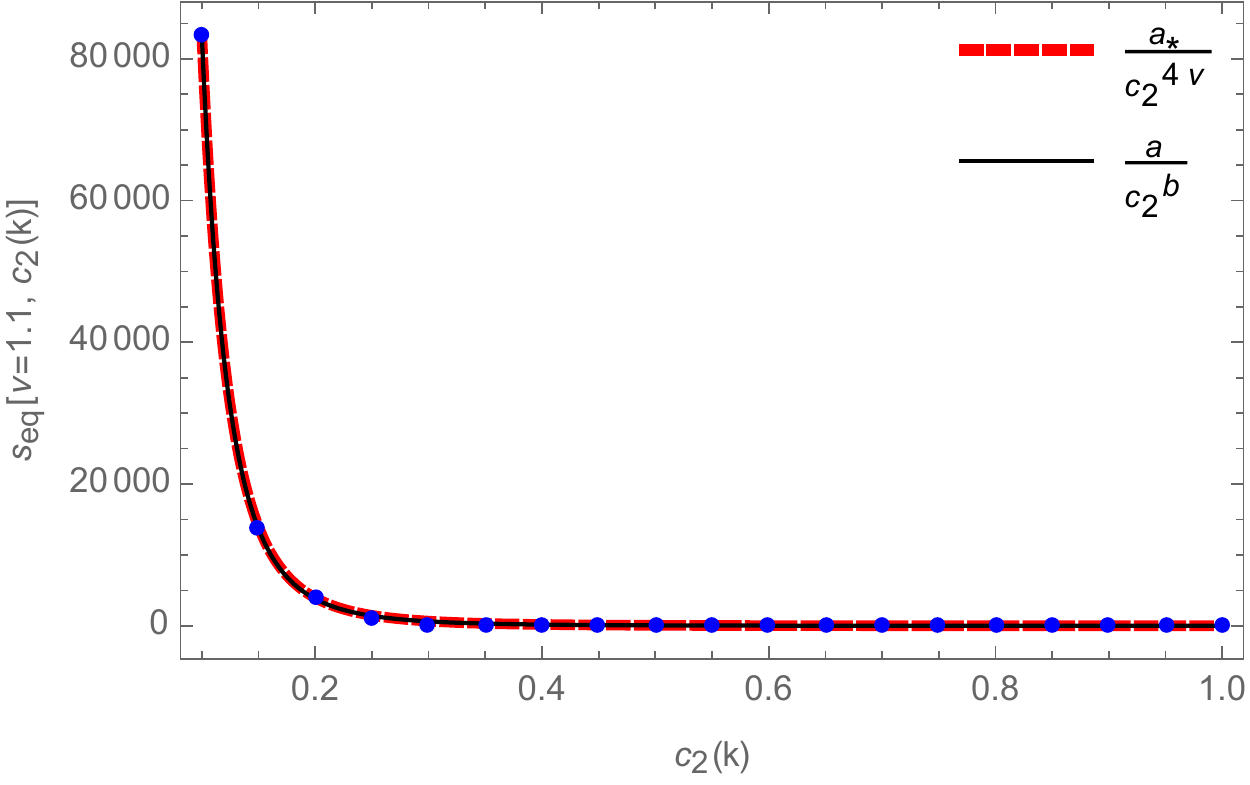}
  \end{subfigure}
  \begin{subfigure}[b]{0.40\textwidth}
    \includegraphics[width=\textwidth]{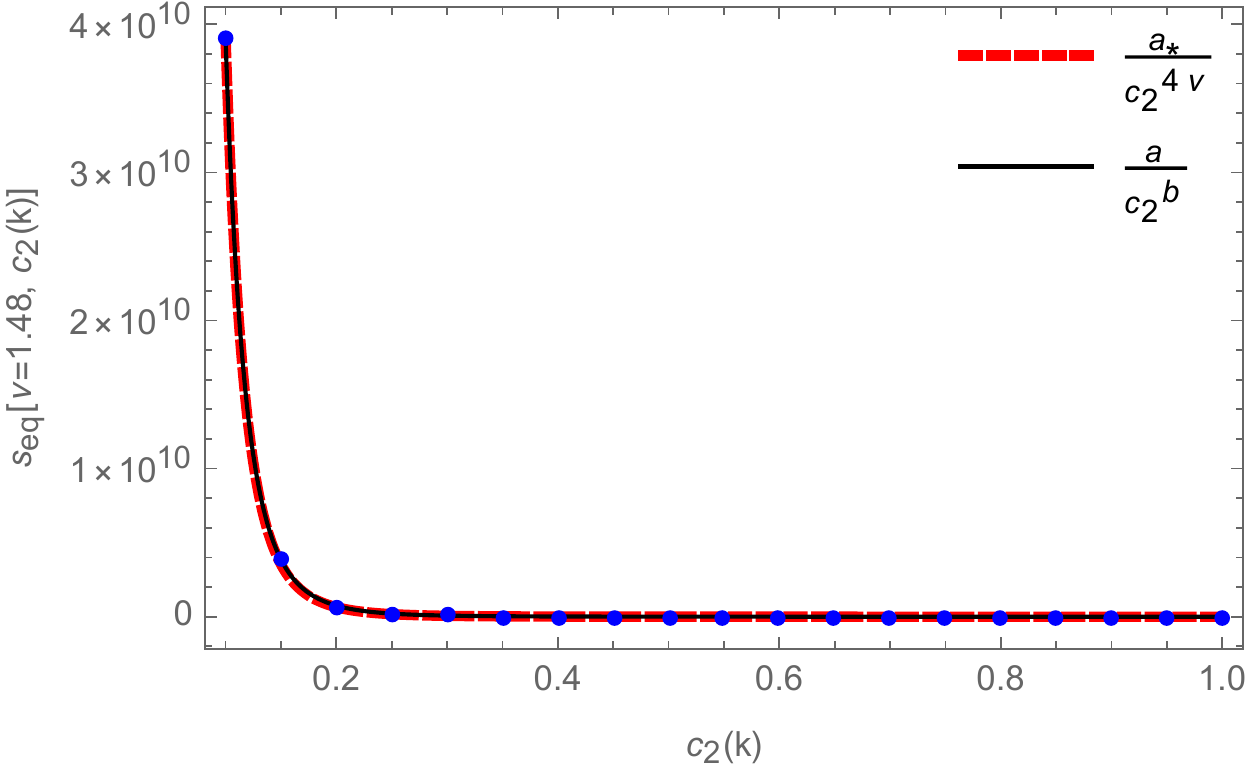}
  \end{subfigure}
  \caption{Numerical results and fitting functions of $s_{eq}[\nu, c_2(k)]$ for $\nu=\{0.4,\, 0.8, \, 1.1, \, 1.48\}$. The plot corresponding to $\nu=1.4$ can be found in the left panel of Fig. \ref{fig:seq14}.}
  \label{fig:seq generic nu}
\end{figure}

\begin{figure}[!ht]
\centering
\captionsetup[subfigure]{justification=centering}
  \begin{subfigure}[b]{0.40\textwidth}
    \includegraphics[width=\textwidth]{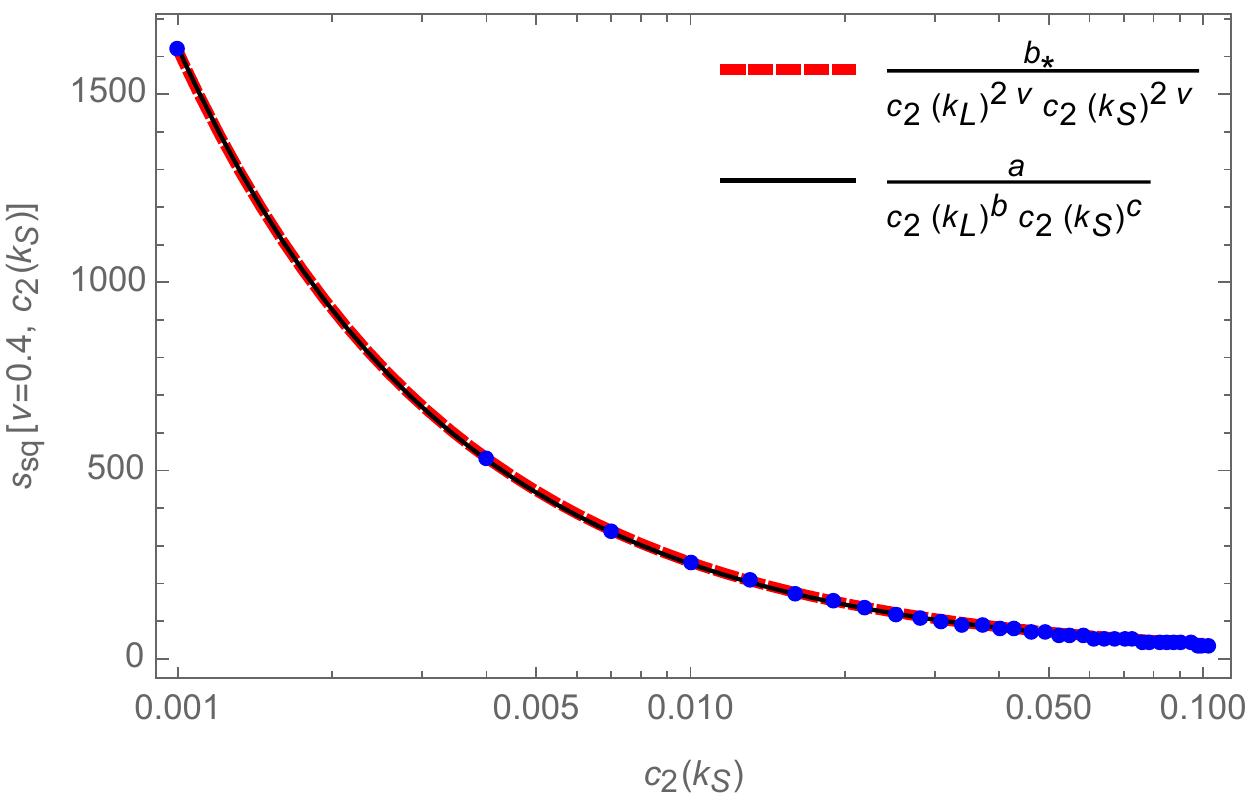}
  \end{subfigure}
  \begin{subfigure}[b]{0.40\textwidth}
    \includegraphics[width=\textwidth]{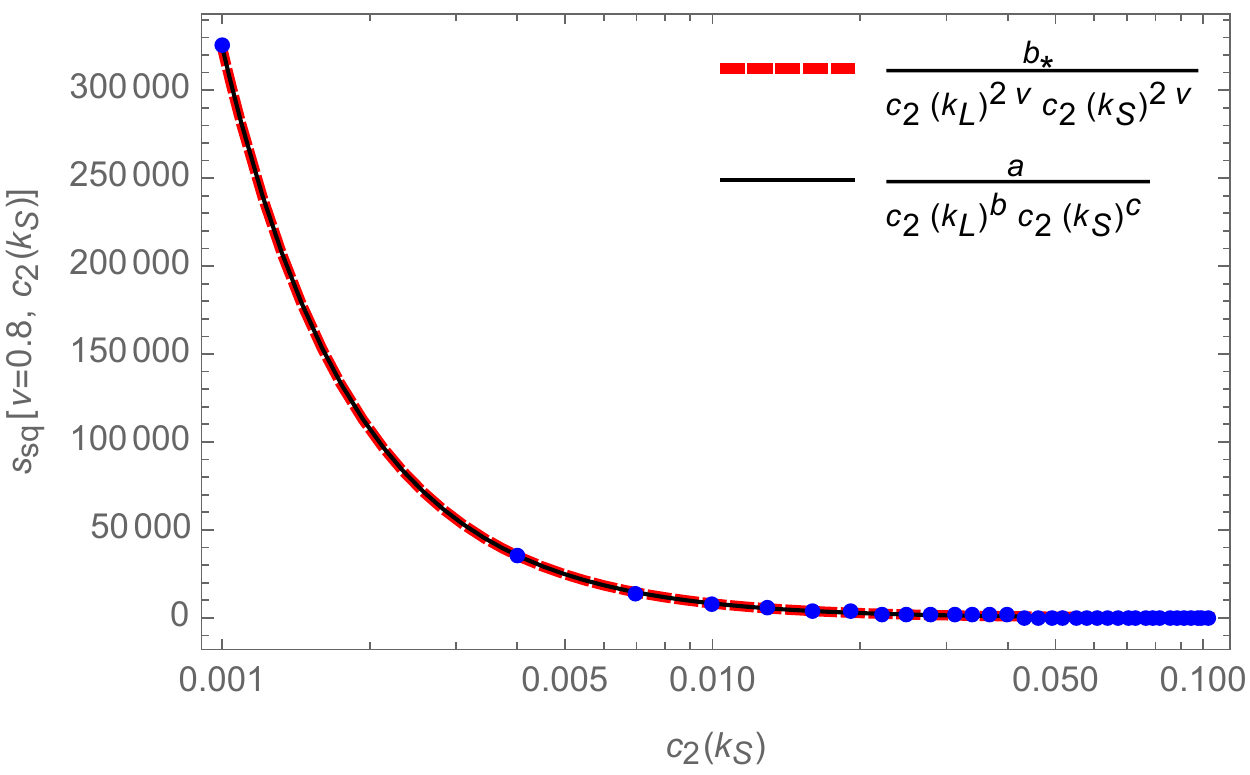}
  \end{subfigure}
  \begin{subfigure}[b]{0.40\textwidth}
    \includegraphics[width=\textwidth]{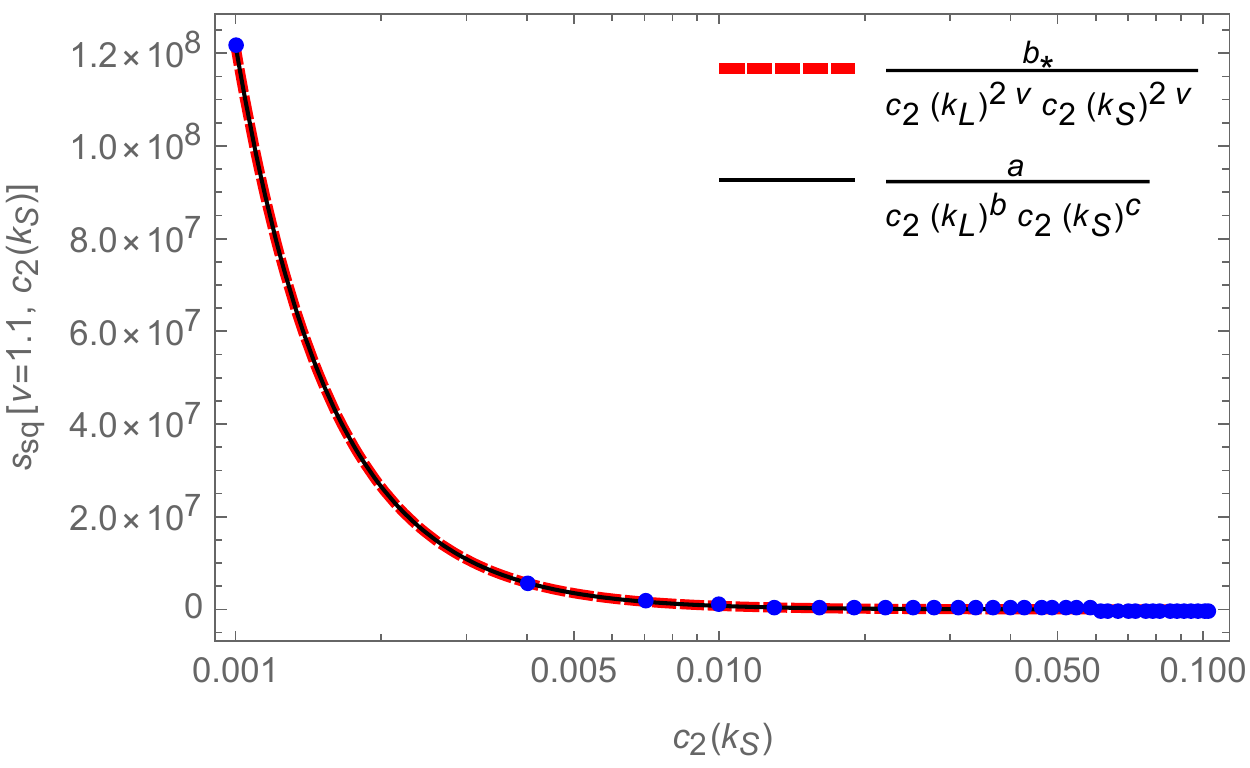}
  \end{subfigure}
  \begin{subfigure}[b]{0.40\textwidth}
    \includegraphics[width=\textwidth]{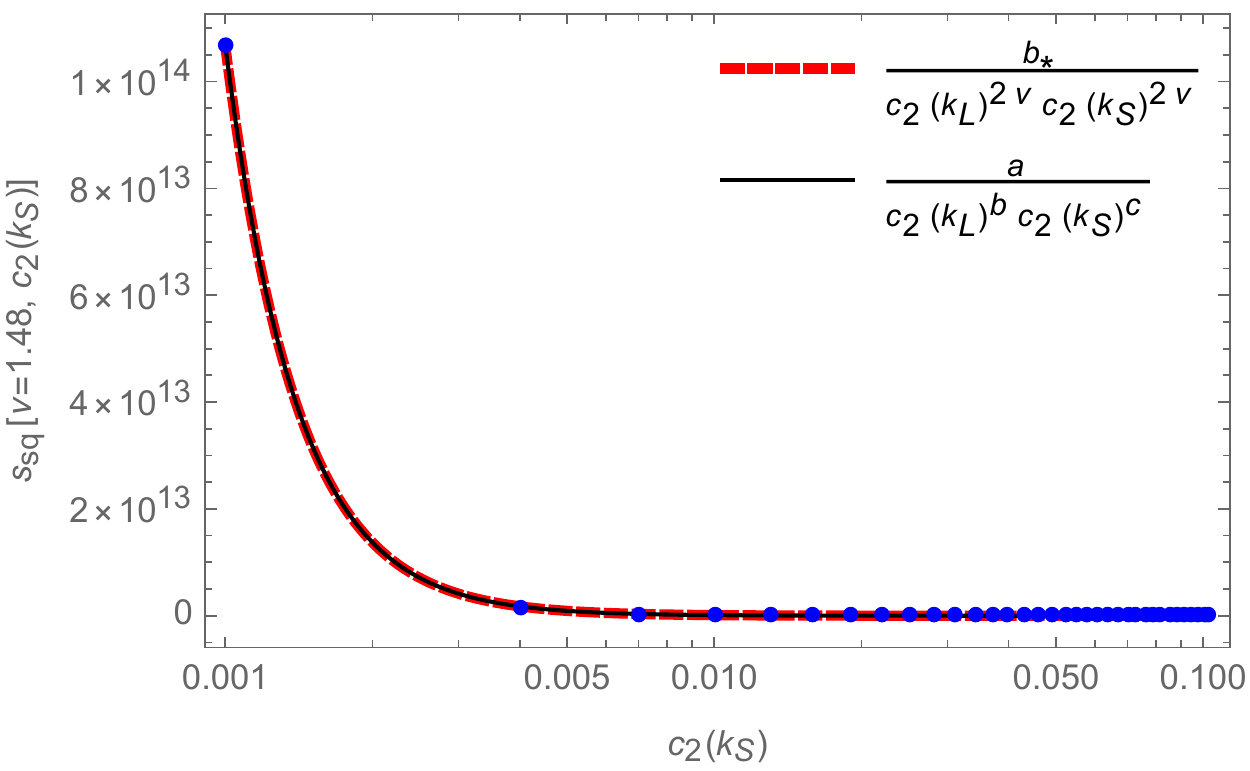}
  \end{subfigure}
  \caption{Numerical results and fitting functions of $s_{sq}[\nu,\, c_2(k_L),\, c_2(k_S)]$ for $\nu=\{0.4,\, 0.8, \, 1.1, \,1.48\}$. In each plot, the sound speed on large scales has been fixed, $c_2(k_L)=0.346$. The plot corresponding to $\nu=1.4$ can be found in the left panel of Fig. \ref{fig:ssq nu 1.4 main}.}
  \label{fig:ssq generic nu}
\end{figure}

While in Section \ref{sec:calculation} our main focus was on the case $\nu=1.4$, we report here some of our findings for the numerical computation of $s_{eq}(\nu)$ and $s_{sq}(\nu)$ for the mass values $\nu=\{0.4,\, 0.8, \, 1.1, \,1.48\}$.
\\ The results in the equilateral and squeezed configurations are displayed in Figs.~\ref{fig:seq generic nu} and \ref{fig:ssq generic nu} respectively. In particular, for each case analysed we fit the numerical values with the power law in Eqs.\eqref{ansatz power law} and \eqref{ansatz power law sq} for the equilateral and squeezed configuration. The fitting functions are plotted with a red dashed line, while the numerical results are represented with blue dots. For completeness, we include also a fit with generic power laws, i.e. leaving the power of $c_2(k)$ free, 
\begin{gather}
    s_{eq}[\nu, c_2(k)]=\frac{a}{c_2(k)^b} \;,\\
     s_{sq} [\nu, c_2(k_L), c_2(k_S)]= \frac{a}{c_2(k_L)^{b} c_2(k_S)^{c} } \;,
\end{gather}
which are plotted in Figs.~\ref{fig:seq generic nu} and \ref{fig:ssq generic nu} with a black continuous line. {The fitting functions Eqs.\eqref{ansatz power law} and \eqref{ansatz power law sq} work better and better towards smaller values of the spin-2 mass ($\nu\rightarrow 3/2$). In the equilateral configuration, the overlap is slightly worse for heavier masses ($\nu\rightarrow 0$).} This must be considered in light of the fact that numerical results for small $\nu$ should not be used for strict quantitative conclusions, as already pointed out in \cite{Chen:2009zp}. In Table \ref{tab:table astar and bstar}, we list the fitted values of $a_\star$ and $b_\star$, defined in Eqs.\eqref{ansatz power law} and \eqref{ansatz power law sq} respectively.

\begin{table}[h]
    \centering
   \begin{tabular}{ |c|c||c|c| }
 \hline
 $\nu$ & $m_\sigma/H$ & $a_\star$& $b_\star$ \\
 \hline
 \hline
0.4& 1.44& 2.7 & 2.8 \\
  \hline
0.8& 1.27 &1.6 &  0.9 \\
  \hline
1.1&  1.02&3.3 & 2.9\\
  \hline
1.4& 0.54&324.4 & 482.8  \\
  \hline
1.48&  0.24&46 876.3 & 19 545.5\\
 \hline
\end{tabular}
    \caption{Values of the fit parameters $a_\star$ and $b_\star$ introduced in Eqs.\eqref{ansatz power law} and \eqref{ansatz power law sq} respectively, obtained for different mass values, $\nu=\sqrt{9/4-(m_\sigma/H)^2}$.}
    \label{tab:table astar and bstar}
\end{table}

\section{Additional details on the squeezed bispectrum}
\label{appendix squeezed}
We report here on the squeezed bispectrum computation, showing how Eq.\eqref{Bsigma squeezed} has been obtained and why the leading contributions come from the A and B terms as spelled out in Eqs.\eqref{FA}-\eqref{FB}, whereas the other permutations and the C term \eqref{FC} are subleading. 
\\ Let us start with the A term, Eq.\eqref{FA}, and take the squeezed limit $k_3\equiv k_L \ll k_1 \sim k_2 \equiv k_S$. For practical purposes, let us consider the large scale to be around CMB scale, $k_L\sim 10^{-2}\mbox{Mpc}^{-1}$, and the small scale to be located for example at LISA scale, $k_S\sim 10^{12}\mbox{Mpc}^{-1}$. Upon the change of variable $y_4\equiv (k_L/k_S)x_4$, Eq.\eqref{FA} can be rewritten as
\begin{equation}
\label{FA sq stage 1}
\begin{split}
 \mathcal{M}_A & (\nu, k_S, k_L) = \Big(\frac{k_S}{k_L}\Big)^{1/2} \int_{-\infty}^{0} dx_1 \;\int_{-\infty}^{x_1} dx_2 \;\int_{-\infty}^{x_2} dx_3 \;\int_{-\infty}^{k_L/k_S\, x_3} dy_4  \sqrt{\frac{x_2}{x_1 x_3 y_4}}\; \sin{(-x_1)} \\
 & \Im\Big[H_\nu^{(1)}(-c_2(k_S) x_1) H_\nu^{(2)}(-c_2(k_S) x_2)\Big] \;
 \Im \Big[\mbox{e}^{-i y_4} H_\nu^{(1)}(-c_2(k_L) y_4) H_\nu^{(2)}(-c_2(k_L) \frac{k_L}{k_S} x_2)\Big] \\
 & \Im\Big[ \mbox{e}^{i x_3 } H_\nu^{(1)}(-c_2(k_S)  x_2) H_\nu^{(2)}(-c_2(k_S) x_3)\Big] \;.
\end{split}
\end{equation}
The Hankel function in the last line, $H_\nu^{(2)}(-c_2(k_S) x_3)$, oscillates and, as a result, suppresses the integral for $c_2(k_S)|x_3| \gg 1$. On small scales the sound speed is of order $10^{-3}$ (see left panel of Fig.~\ref{fig:c2(k)eq}), therefore only values $|x_3|\ll 10^3$ are relevant for the integral computation. As a consequence, the upper limit of the integral in $y_4$ is effectively zero for the reference scales considered. 
\\ Moreover, by looking at the Hankel function $ H_\nu^{(1)}(-c_2(k_S)  x_2)$, one can infer that only values $|x_2|\ll 10^3$ contribute to the integral. Therefore, the Hankel function $H_\nu^{(2)}(-c_2(k_L)\, k_L/k_S \, x_2)$ can be approximated in the small argument limit, $ H^{(2)}_\nu (x)\rightarrow i \, 2^\nu \Gamma (\nu) x^{-\nu}/\pi$. Indeed, on large scales the sound speed is of order $0.1$ (right panel of Fig.~\ref{fig:c2(k)eq}), so the argument of the Hankel is very small, $\mathcal{O}(10^{-12})$. As a result of these approximations, Eq.\eqref{FA sq stage 1} reduces to 
\begin{equation}
\label{FA sq}
\begin{split}
 \mathcal{M}_A & (\nu, k_S, k_L)= \frac{2^\nu \Gamma(\nu)}{\pi c_2(k_L)^\nu}  \Big(\frac{k_S}{k_L}\Big)^{1/2+\nu} \int_{-\infty}^{0} dx_1 \;\int_{-\infty}^{x_1} dx_2 \;\int_{-\infty}^{x_2} dx_3  \times \\
 & (-x_2)^{1/2-\nu} (-x_1)^{-1/2}(-x_3)^{-1/2}\; \sin{(-x_1)} \Im\Big[H_\nu^{(1)}(-c_2(k_S) x_1) H_\nu^{(2)}(-c_2(k_S) x_2)\Big] \\
 & \Im\Big[ \mbox{e}^{i x_3 } H_\nu^{(1)}(-c_2(k_S) x_2) H_\nu^{(2)}(-c_2(k_S) x_3)\Big] \times \int_{-\infty}^{0} dy_4 (-y_4)^{-1/2}\Re \Big[\mbox{e}^{-i y_4} H_\nu^{(1)}(-c_2(k_L) y_4) \Big] \;.
\end{split}
\end{equation}
A similar analysis can be performed for the B term in Eq.\eqref{FB}, to give 
\begin{equation}
\label{FB sq}
\begin{split}
 \mathcal{M}_B & (\nu, k_S, k_L)= \frac{2^\nu \Gamma(\nu)}{\pi c_2(k_L)^\nu}  \Big(\frac{k_S}{k_L}\Big)^{1/2+\nu} \int_{-\infty}^{0} dx_1 \;\int_{-\infty}^{x_1} dx_2 \;\int_{-\infty}^{x_2} dx_3  \; (-x_3)^{1/2-\nu} (-x_1)^{-1/2}(-x_2)^{-1/2}\\
 &\sin{(-x_1)} \sin{(-x_2)} \Im\Big[H_\nu^{(1)}(-c_2(k_S) x_3) H_\nu^{(1)}(-c_2(k_S) x_3) H_\nu^{(2)}(-c_2(k_S) x_1)H_\nu^{(2)}(-c_2(k_S) x_2) \Big] \\
 & \times \int_{-\infty}^{0} dy_4 (-y_4)^{-1/2}\Re \Big[\mbox{e}^{-i y_4} H_\nu^{(1)}(-c_2(k_L) y_4) \Big] \;.
\end{split}
\end{equation}
 \begin{table}[]
\centering
\begin{tabular}{ |c|c|c|c| } 
\hline
Term & Permutation & Scaling \\
\hline \hline
\multirow{3}{*}{A} & as spelled in Eq.\eqref{FA sq} & $k_S^{-9/2+\nu} k_L^{-3/2-\nu}$ \\ [1ex]

& $k_3\leftrightarrow k_1$ & $k_S^{-6}$ \\[1ex]

& $k_3\leftrightarrow k_2$  & $k_S^{-5} k_L^{-1}$ and $k_S^{-6+2\nu} k_L^{-2\nu}$ \\ [1ex]
\hline
\multirow{3}{*}{B} & as spelled in Eq.\eqref{FB sq} & $k_S^{-9/2+\nu} k_L^{-3/2-\nu}$ \\ [1ex]

& $k_3\leftrightarrow k_1$ & $k_S^{-6+2\nu} k_L^{-2\nu}$ \\[1ex]

& $k_3\leftrightarrow k_2$  & $k_S^{-6+2\nu} k_L^{-2\nu}$ \\ [1ex]
\hline
\multirow{3}{*}{C} & as spelled in Eq.\eqref{FC} & $k_S^{-6+2\nu} k_L^{-2\nu}$ \\ [1ex]

& $k_3\leftrightarrow k_1$ & $k_S^{-6+2\nu} k_L^{-2\nu}$ \\[1ex]

& $k_3\leftrightarrow k_2$  & $k_S^{-6+2\nu} k_L^{-2\nu}$ \\ [1ex]
\hline
\end{tabular}
\caption{Scaling behavior of the different contributions. For $A$ ($k_3 \leftrightarrow k_2$) the scaling is different depending on the value of the mass: the first one is valid for $\nu<1/2$ and the second for $1/2<\nu<3/2$.}
\label{tab:scaling table}
\end{table}

\begin{figure}
\centering
\captionsetup[subfigure]{justification=centering}
   \begin{subfigure}[b]{0.49\textwidth}
    \includegraphics[width=\textwidth]{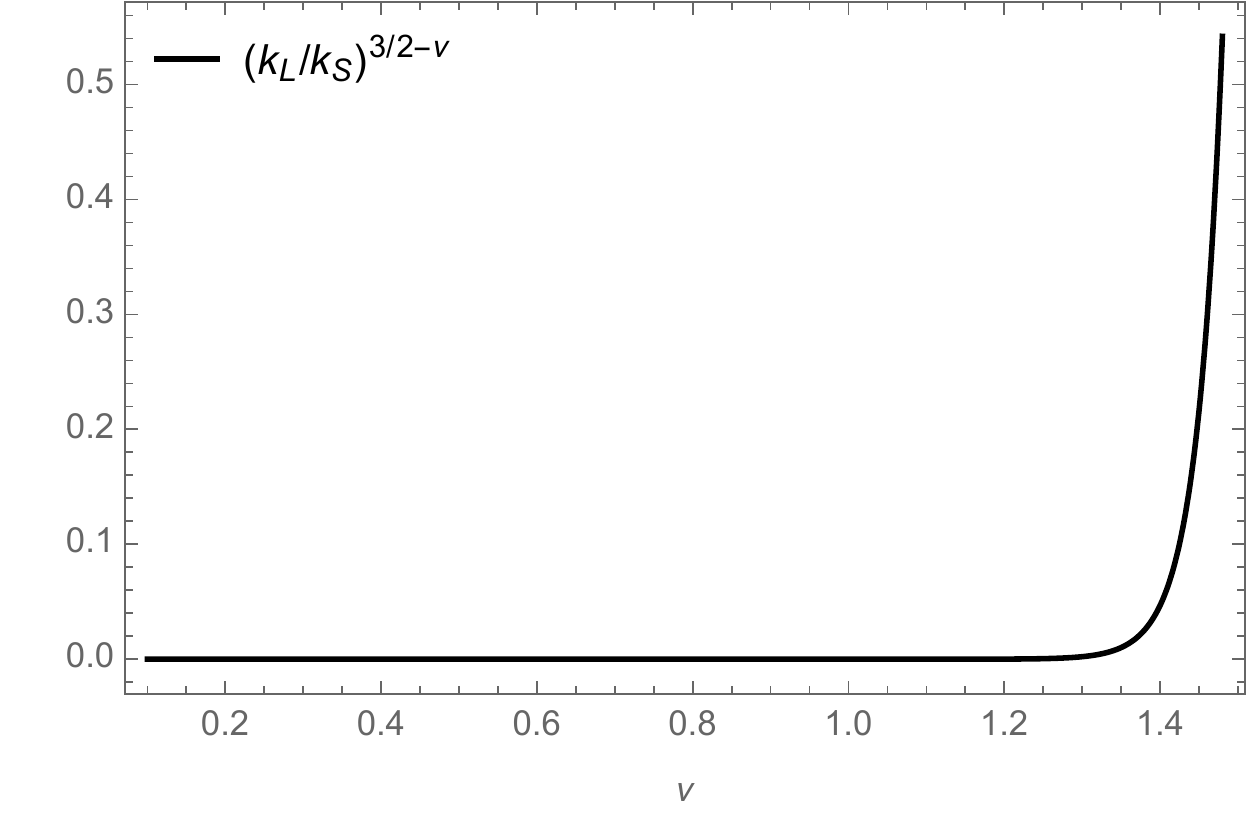}

  \end{subfigure}
  \begin{subfigure}[b]{0.49\textwidth}
    \includegraphics[width=\textwidth]{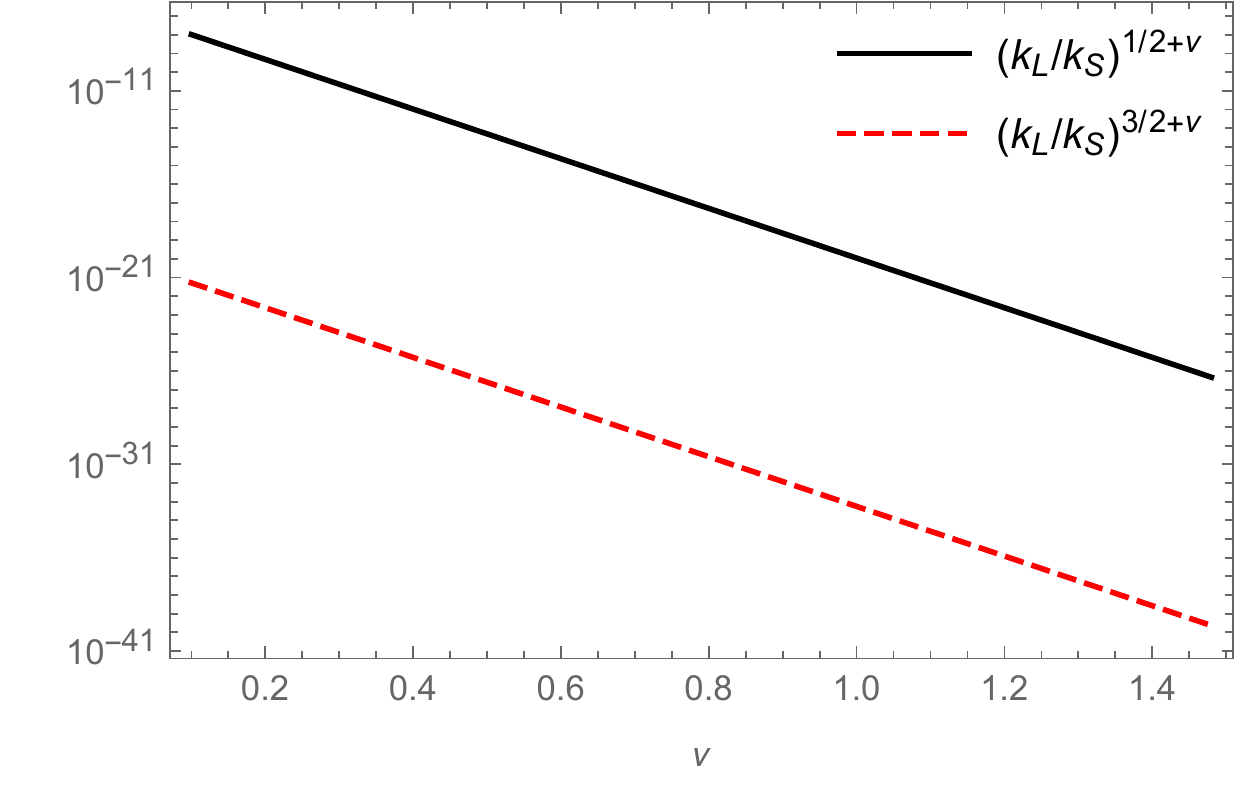}

  \end{subfigure}
  \caption{Plots representing the functions in Eqs.\eqref{scaling1} (left) and \eqref{scaling2}-\eqref{scaling3} (right) with respect to the mass $\nu$, with $k_S=0.05\,\mbox{Mpc}^{-1}$ and $k_L=10^{12}\,\mbox{Mpc}^{-1}$.}
  \label{fig:scaling beha}
\end{figure}
\noi The sum of these two contributions results in Eq.\eqref{Bsigma squeezed}, where the overall explicit\footnote{Note that in Eq.\eqref{ssq}, as well as for each term in Eq.\eqref{Bispectrum all terms}, there is also an additional hidden scaling due to the scale dependence of the sound speed $c_2(k)$ (see Section \ref{sec:calculation}). For completeness, we have explicitly numerically evaluated all the contributions in Eq.\eqref{Bispectrum all terms} for masses $\nu=\{0.4,\,0.8,\, 1.1,\, 1.4,\, 1.48\}$ with $\{c_{2\,in}=1,\, k_L=0.05\, \mbox{Mpc}^{-1}, \, k_S=10^{12}\, \mbox{Mpc}^{-1}\}$ and confirmed the conclusions described in the main text: looking at the explicit scaling of each term is enough to establish whether it contributes or not.} scaling behavior is 
\begin{equation}
\label{main scaling}
    \frac{1}{k_L^{9/2-\nu}k_S^{3/2+\nu}}    \;.
\end{equation}
\\ We proceed in a similar fashion to study the squeezed limit of the C term, Eq.\eqref{FC}, and all the permutations in Eq.\eqref{Bispectrum all terms} (here we refer to the permutations $k_3 \leftrightarrow k_2$ and $k_3 \leftrightarrow k_1$, while $k_1 \leftrightarrow k_2$ contributes with a factor $2$). The resulting scalings are listed in Table \ref{tab:scaling table}. The contribution of each term relative to the those spelled out in Eqs.\eqref{FA sq}-\eqref{FB sq} is classified by looking at the ratio of the scaling with respect to that in \eqref{main scaling}. 
For $k_L=0.05 \,\mbox{Mpc}^{-1}$ and $k_S= 10^{12}\,\mbox{Mpc}^{-1}$, we plot on the left panel of Fig.~\ref{fig:scaling beha} the function 
\begin{gather}
\label{scaling1}
    \frac{1}{k_S^{6-2\nu} k_L^{2\nu}}\Big/ \frac{1}{k_S^{9/2-\nu} k_L^{3/2+\nu}}= \Big(\frac{k_L}{k_S} \Big)^{3/2-\nu}
\end{gather}
and on the right panel the functions 
\begin{gather}
\label{scaling2}
    \frac{1}{k_S^{6}}\Big/ \frac{1}{k_S^{9/2-\nu} k_L^{3/2+\nu}}= \Big(\frac{k_L}{k_S} \Big)^{3/2+\nu} \\
\label{scaling3}
     \frac{1}{k_S^{5}k_L}\Big/ \frac{1}{k_S^{9/2-\nu} k_L^{3/2+\nu}}= \Big(\frac{k_L}{k_S} \Big)^{1/2+\nu} \;.
\end{gather}
We conclude that the $A$ term ($k_3 \leftrightarrow k_1$) is always subleading for all masses. For $\nu<1/2$ also the $k_3 \leftrightarrow k_2$ permutation can be safely neglected. For $1/2<\nu<3/2$ the $k_3 \leftrightarrow k_2$ permutation of $A$ can be safely neglected for most of the mass values, whereas must be considered for $\nu\rightarrow3/2$ as the scaling is no more suppressed with respect to that in \eqref{main scaling} (see left panel Fig.~\ref{fig:scaling beha}). The same consideration holds for $B$($k_3 \leftrightarrow k_1$), $B$($k_3 \leftrightarrow k_2$) and the $C$ term.

\end{appendices}

\bibliography{main} 

\bibliographystyle{JHEP}

\end{document}